\documentclass[%
 reprint,
 amsmath,amssymb,
 aps,
]{revtex4-2}
\pdfoutput=1 
\usepackage{graphicx}
\usepackage{dcolumn}
\usepackage{bm}
\usepackage{xcolor}
\usepackage[utf8]{inputenc}
\usepackage[T1]{fontenc}
\usepackage{soul}
\usepackage[nolist,nohyperlinks]{acronym}

\begin{document}

\preprint{APS/123-QED}
\title{Principles for single-pixel terahertz imaging based on the engineering of illuminating and collecting nonparaxial diffractive optics}

\author{Sergej Orlov}%
\author{Rusnė Ivaškevičiūtė-Povilauskienė}
\author{Karolis Mundrys}
\author{Paulius Kizevičius}%
\author{Ernestas Nacius}%
\author{Domas Jokubauskis}
\affiliation{Center for Physical Sciences and Technology (FTMC), Saulėtekio Ave.~3, LT-10257 Vilnius, Lithuania \\}%

\author{Kęstutis Ikamas}%
\author{Alvydas Lisauskas}%
\affiliation{Physics Department, Vilnius University, Saulėtekio Ave. 3, LT-10257 Vilnius, Lithuania \\}%

\author{Linas Minkevičius}%
\author{Gintaras Valušis}%
\affiliation{Center for Physical Sciences and Technology (FTMC), Saulėtekio Ave.~3, LT-10257 Vilnius, Lithuania\\}%


\date{\today}

\begin{abstract}

The art of light engineering unveils a world of possibilities through the meticulous manipulation of photonic properties such as intensity, phase, and polarization. This technique offers an excellent opportunity to enhance photonic signal quality, mitigate unwanted interference sources, and finely sculpt the spatial characteristics of light. The precise control over these optical properties under various conditions finds application in fields spanning communication, light-matter interactions, laser direct writing, and imaging, enriching our technological landscape.
Meeting the specific demands of diverse photonic applications presents a formidable challenge, hinging on factors like wavelength and radiation power. Within this context, the terahertz (THz) range, nestled between microwaves and infrared light, stands out for its remarkable ability to propagate with minimal losses in numerous dielectric materials and compounds. This unique feature allows noninvasive control and inspection, making THz imaging a powerful tool for nondestructive assessment of dielectric materials, medical diagnostics, chemical identification, security checks, and industrial monitoring.
In this study, we embark on a journey to establish a rational framework for the design and assembly of nonparaxial THz imaging systems. Our focus centers on a lensless photonic system composed solely of flat-silicon diffractive optics. These elements include the high-resistivity silicon-based nonparaxial Fresnel zone plate, the Fibonacci lens, the Bessel axicon, and the Airy zone plate, all meticulously crafted using laser ablation technology. A systematic exploration of these flat elements in various combinations sheds light on their strengths and weaknesses.
Our endeavor extends to the practical application of these optical components, where they illuminate samples and capture the light scattered from these raster-scanned samples using single-pixel detectors. Through a comprehensive examination, we evaluate imaging systems across diverse metrics that include contrast, resolution, depth of field and focus. This multifaceted approach allows us to distill rational design principles for the optimal assembly of THz imaging setups.
The findings of this research chart an exciting course toward the development of compact, user-friendly THz imaging systems where sensors and passive optical elements seamlessly integrate into a single chip. These innovations not only enhance capabilities in THz imaging but also pave the way for novel applications, ushering in a new encouraging era of advanced THz photonic technology.
\end{abstract}

\maketitle

\begin{acronym}
\acro{DOF}{depth of field}
\acro{MTF}{modulation transfer function}
\acro{THz}{terahertz}
\acro{CW}{continuous wave}
\acro{FWHM}{full width at half-maximum}
\acro{ZP}{zone plate}
\acro{2ZP}{second zone plate}
\acro{SLM}{spatial light modulator}
\acro{SNR}{signal-to-noise ratio}
\acro{DoFo}{depth of focus}
\end{acronym}

\section{\label{sec:level1}Introduction\protect\\}

The engineering of electromagnetic fields to create photonic structures, as required by a number of particular applications \cite{rubinsztein2016roadmap} is attracting attention due to its venerable nature and the beautiful physics behind it. Therefore, new possibilities for generating and detecting structured light can play an essential role in extending its application areas \cite{angelsky2020structured}. The concept of beam engineering was successful in extending the development in photonic communications \cite{yuan2017beam,zhu2020airy}, laser microfabrication \cite{baltrukonis2021high,vslevas2022azimuthally}, imaging improvements \cite{geng2011structured,bitman2012improving,minkevicius2019bessel}, tomography \cite{wang2020airy}, light sheet microscopy \cite{vettenburg2014light}, and microscopy within a wide range of the electromagnetic spectrum, covering frequencies from visible to \ac{THz}  \cite{zhang2021enhanced}. Structured light illumination and optimal placement of the optical elements in its collection can lead to high resolution, sharpness, and contrast in images, which is one of the main objectives of imaging theory. 

Terahertz imaging stands out as a powerful technique for nondestructive inspection in different types of applications, including materials science, biomedical examination, or security checks \cite{Castro-Camus2022,yan2022thz, Valusis2021}. However, practical implementation of compact \ac{THz} imaging systems in a real operational environment still faces significant obstacles due to low \ac{THz} emitter powers, reliability of sensitive \ac{THz} detectors, and effective designs in planar optical component technology, as well as a lack of optimal solutions that allow one to avoid the need for precise optical alignment \cite{Valusis2021}. In particular, as the most investigated objects are rather bulky and the image quality is strikingly sensitive to the position of the sample with respect to passive optical elements such as mirrors or lenses, the obtained images can suffer from low contrast, can be blurred, or can hardly be resolved.
 
Flat optical components \cite{banerji2019imaging,engelberg2020advantages,reshef2021optic,Advances_meta-holography} are a convenient replacement for bulky optical systems. The thickness and weight of the optical elements can be greatly reduced by exploiting the phenomenon of diffraction \cite{yu2014flat, genevet2017recent,headland2018tutorial,banerji2019imaging}. The spatial arrangement of sub- or near-wavelength thickness elements to produce phase shifts in a passing optical ray leads to the desired constructive interference of the transmitted waves at the given point of observation. Implementing this concept involves binary or multilevel diffractive elements where the phase delay is $(n-1)t$ for the material of refractive index $n$ and the local thickness $t$ of the substructure. Another implementation is based on the concept of metasurfaces, where diffractive gratings are realized by local changes in the geometric phase of the structure consisting of individual metaatoms \cite{Visible_metagratings, Vectorial_compound_metapixels}.

This advancement has resulted in a greatly reduced mass and thickness of Fresnel lenses \cite{banerji2019imaging, Siemion2019}, top-hat converters \cite{gotovski2022investigation}, and axicons \cite{vasara1989realization,minkevicius2019bessel}. An axicon became especially interesting, as it has enabled the generation of the so-called non-diffracting Bessel beam \cite{ren2021non,vasara1989realization,wei2015generation}, which can be further engineered using a flat geometrical phase element \cite{gotovski2021generation}. Elliptic Mathieu \cite{gutierrez2000alternative}, parabolic Weber \cite{bandres2004parabolic}, and self-accelerating Airy beams \cite{siviloglou2007observation, Broadband_generation_visible} is yet another member of the attractive family of non-diffracting beams. It is worth noting that flat optical elements can handle all those nondiffracting beams efficiently, for instance, to generate Airy beams under various conditions \cite{guo2020airy} within a wide range of wavelengths extending from optical \cite{Broadband_generation_visible,wen2021all} to \ac{THz} \cite{cheng2021achromatic}. 

Laser ablation technology is a convenient tool in the fabrication of silicon diffractive zone plates \cite{Minkevicius2017}. Using their advantage in maintaining the same technological conditions, different elements were designed and produced to shape the \ac{THz} radiation of the Gaussian modes to the Fibonacci structure \cite{Minkevicius2018}, prepare the Bessel beam \cite{minkevicius2019bessel}, and generate structured \ac{THz} light in the form of the Airy beam \cite{ivaskeviciute2022} in the \ac{THz} imaging experiments. It was revealed that beam engineering, structuring the \ac{THz} illumination, has resulted in advanced imaging capabilities, increased resolution, contrast, and sensitivity to material parameters of thin samples \cite{ivaskeviciute2022}.

Contemporary cameras and cell phones capture images with a single shot using detector arrays and lenses with millions of pixels. Although increasing the number of pixels beyond 20 million seems unnecessary and results in data storage issues, an alternative method of image retrieval involves using a single-pixel detector.  This technique has been in use for over a century and is still commonly used in non-visible spectrum applications where detector arrays of certain wavelengths are either expensive or unavailable. 

Single-pixel imaging is an innovative approach to image reconstruction that involves the utilization of a compact small and sensitive detector. Traditionally, this technique involves quantifying the degree of overlap in the scene with various masks using a solitary element detector \cite{pittman1995optical} that subsequently integrates these measurements with the mask information \cite{shapiro2008computational}. The viability of this approach has been rigorously substantiated through theoretical \cite{duarte2008single} and experimental validation \cite{edgar2019principles}.

In recent years, ghost imaging has rekindled enthusiasm for research within the realm of single-pixel imaging architectures, following its inaugural experimental demonstration \cite{pittman1995optical,shapiro2008computational}. Ghost imaging is sometimes classified as a computational imaging technique, which means that data acquired by a ghost imaging system requires processing through computational algorithms to manifest as a conventional image. Both the single-pixel imaging community and the ghost imaging community soon realized that the two imaging architectures are essentially the same optically \cite{qiu2020comprehensive}. This intriguing convergence between the single-pixel imaging and ghost imaging communities has revealed the fundamental optical equivalence of these imaging architectures.  

In a classical sense, single-pixel imaging involves the integration of \acp{SLM} on the focal plane of the camera lens, which facilitates the modulation of scene imagery using various masks prior to measuring light intensities with a single-pixel detector. Ghost imaging, on the contrary, employs diverse structured light distributions, created by 
\acp{SLM}, to illuminate the scene, subsequently capturing the reflected or transmitted light intensities. In all of these methods, instead of capturing individual pixel samples from the observed scene, the inner products between the scene and a collection of test functions are assessed. In particular, the use of random illumination or light collection functions plays a pivotal role in this approach, rendering each measurement a stochastic sum of pixel values spanning the entire image.

These approaches are most popular within the community working with single-pixel imaging, as they use the \ac{SLM} either to spatially modulate the illuminating or collecting schemes. An additional degree of classification is image scanning techniques: three different scanning techniques are actively used \cite{duarte2008single}: basis scan, raster scan, and compressive sampling. In the work described here, we employ a raster scan technique.

 
Direct implementation of \ac{THz} imaging requires enhanced functionality, compact solutions in the design of optical setup, and greater convenience of use. The preferable compact solution in the optical setup of \ac{THz} imaging can be, for example, the integration of passive optical elements with detectors on a single chip \cite{Minkevicius2014E}. The image recording system is compact, integrated into the chip, displays enhanced sensitivity, and is free of optical alignment. However, from the point of view of optical design, it presents challenges. In this situation, \ac{THz} illumination incident on the object is very close to the optical axis, the focus is very sharp, in the range of the substrate thickness, thus inducing the high aperture, and the beam waist becomes comparable with the wavelength of the \ac{THz} radiation. It means that conventional paraxial optics are not valid here and for a correct description, one needs to use the nonparaxial illumination approach.
 
This task is not trivial due to the absence of image formation rules for nonparaxial single-pixel imaging \cite{duarte2008single,edgar2019principles,gibson2020single}. Single-shot imaging is well established for lenses containing systems \cite{iizuka2013engineering} -- the formula $z_1^{-1}+z_2^{-1}=f^{-1}$, where the distances $z_1$, $z_2$ are between the object and the lens and between the image and the lens, respectively, and $f$ is the focal length -- gives a determined imaging law, rational design rules for conditions when spatial resolution and contrast are optimal, and the enhancements and aberrations introduced by the axicons and other elements are well understood \cite{arimoto1992imaging,zhai2009extended}. However, in particular, in the case of structured light with respect to the displacement of diffractive optical elements, single-pixel imaging scenarios are not investigated in detail; also, principles of rational design of optical setups for \ac{THz} imaging in the nonparaxial illumination case are still lacking.

In this work, we showcase the importance of structuring the illumination and collection parts of the optical system in a single-pixel raster-scan method. Moreover, we establish principles for silicon diffractive optics - based \ac{THz} light engineering and optical components positioning enabling rational design in \textbf{a compact single-pixel nonparaxial \ac{THz} imaging}. Numerical and experimental investigations in a variety of object illumination–collection schemes involving symmetric and asymmetric combinations of Fresnel zone plates, Fibonacci lenses, the Bessel axicon, and the Airy zone plate allowed us to determine the effect of the different optical configurations on the parameters of the recorded THz images. It is shown that \textbf{in nonparaxial imaging, the optimal modulation transfer condition appears to be decoupled from the condition for the best image irradiance.} The use of different high-resistivity silicon-based diffractive optic elements produced by laser ablation technology under the same technological conditions allowed the quality of the fabrication to remain the same and stable, allowing precise benchmarking of the performance of imaging systems. It is demonstrated that \ac{THz} \textbf{structured light generated in nonparaxial imaging systems \ac{THz} can improve image quality in terms of contrast, resolution, and depth of field, and can be done in lensless compact single-pixel \ac{THz} optical setups containing only diffractive optical elements.} The peripheralities due to \ac{THz} light nonparaxiality in the image quality assessment of single pixels are discussed in detail and compared with those of single-shot imaging recording schemes.


\section{\label{sec:design_principles}Theoretical frame and design principles\protect\\}

We design a flat diffractive photonic element for a rather long wavelength (compared to the characteristic dimensions of the imaged objects) $\lambda$ of \ac{THz} radiation. We consider a set of flat optical elements, a conventional Fresnel zone plate, a Fibonacci lens, an axicon to produce a Bessel beam, and an Airy mask to generate Airy illumination. The complex transmission function $T(\mathbf{r})=\exp \left[ i \Phi (\mathbf{r}) \right]$ of a multilevel phase mask is defined by the introduction of a phase $\Phi (\mathbf{r})$:

\begin{equation}
\Phi (\mathbf{r})=
\frac{2\pi }{N}
\left\lfloor \frac{N \Psi \left(x,y\right)}{2 \pi}
-N \left \lfloor\frac{ \Psi \left(x,y\right)}{2 \pi}\right\rfloor\right\rfloor ,
\label{Eq:Tq}
\end{equation}
where $\Psi \left(x,y\right)$ is a continuous phase profile of the designed element and the brackets $\lfloor~\rfloor$ represent a rounding operation with $N$ being the integer number of levels in the phase mask. The resulting phase profile $\Phi (\mathbf{r})$ can be height encoded using the relation $\lambda \Phi (\mathbf{r}) = 2 \pi (n-1) h (\mathbf{r})$, where $h (\mathbf{r})$ are local near-wavelength variations in surface height.

First, we design a phase zone plate with the phase of the transmission function 
\begin{equation}
\Psi_{\text{ZP}}(\mathbf{r})=\frac{\pi}{f\lambda}\left(x^{2}+y^{2}\right),
\label{eq:Tl}
\end{equation}
where $f=1$~cm is the expected paraxial focus length. Similar designs and their performance are discussed in~\cite{Minkevicius2017, Minkevicius2018, Indrisiunas2019}.

The Fibonacci lens was created using a procedure described in Ref.~\cite{monsoriu2013bifocal} using Fibonacci sequences with $n_{\text{seq}}=7$. The element was a binary element comprising parts of the surface with phases $0$ and $\pi$.

The generation of the Bessel \ac{THz} beam was performed using a linear phase function.

\begin{equation}
\Psi_{\text{B}}(\mathbf{r})=\frac{2 \pi \sin \beta}{\lambda}\sqrt{x^{2}+y^{2}}
\label{eq:Tb}
\end{equation}
where $\beta =0.4$~rad.

To engineer the Airy beam, we devised a cubic phase profile as  

\begin{equation}
    \Psi_{\text{AI}}(\mathbf{r})=a\left(x^{3}-y^{3}\right),
    \label{eq:S}
\end{equation}
($a=\pi \times 10^{7}\text{m}^{-3}$). This design represents a phase mask plate of eight levels ($N=8$) of diameter $20$~mm, which together with a zone plate ($f = 1$~cm) is dedicated to generating an Airy beam in the range of up to $10$~mm.

We recall the Rayleigh-Sommerfeld diffraction integral \cite{born2013principles}.
\begin{equation}
    U(\mathbf{r}_1)=
    \frac{1}{ i\lambda}
    \int_{S_{\text{A}}}
    U_{\text{inc}}(\mathbf{r}_{0})
    T(\mathbf{r}_{0})
    \frac{\text{e}^{\text{i} k\left(\mathbf{r}_{01}\right)}
    }{\lvert  \mathbf{r}_{01} \rvert } 
    \cos \left(\textbf{r}_{01}, \textbf{n}\right)  d S ,
    \label{Somm}
\end{equation}
where $U(\mathbf{r}_1)$ is the electric field in the observation plane, $U_{\text{inc}}(\mathbf{r}_{0})$ is the incident electric field in the diffraction plane, $T(\mathbf{r}_{0})$ is the transmittance of the object. The coordinates of the illumination plane are $\mathbf{r}_{0}=(x_0,y_0,z=0)$ and the coordinates of the observation plane are $\mathbf{r}_{1}=(x_1,y_1,z=z_{o})$. The vector $\mathbf{r}_{01}$ defines the distance between two points in these planes, and $\textbf{n}$ is a normal direction toward the surface of the object. Integration is performed on the surface of the element $S_\text{A}$. Obviously, a propagator here is a spherical point source to numerically simulate the propagation of electromagnetic fields.

\section{\label{sec:fabrication_and_setup}Experimental setup, fabrication of diffractive elements, and experimental verification of their performance\protect\\}

\subsection{\label{sec:fabrication} Experimental setup and fabrication of diffractive optical elements}

\begin{figure}
    \centering
    \includegraphics[width=1\columnwidth]{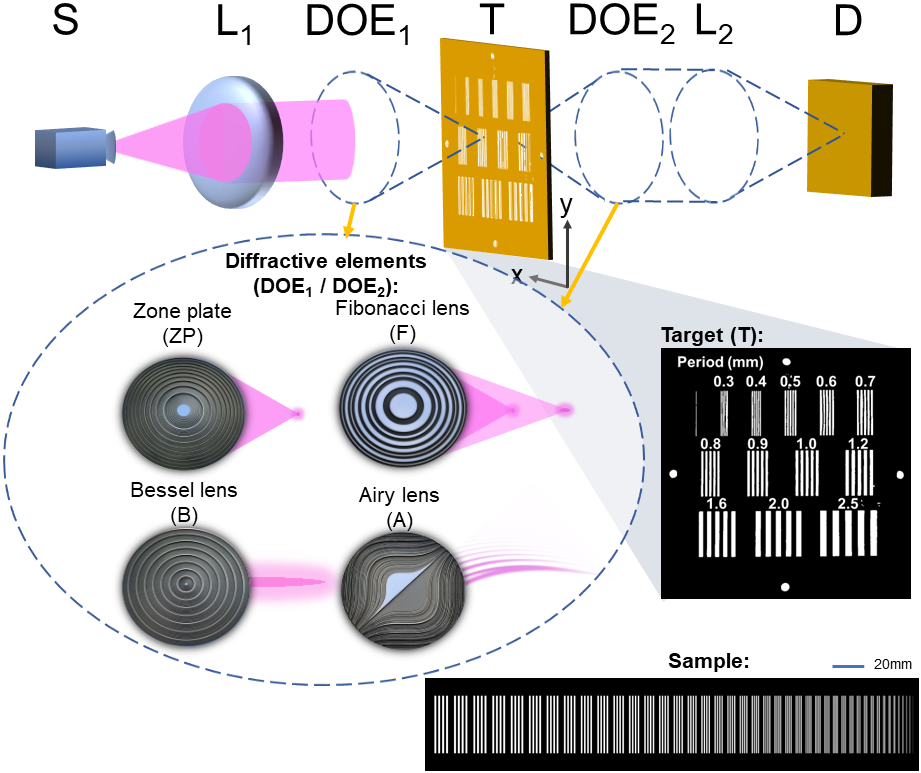}
    \caption{Experimental setup for \ac{THz} imaging. The letter S denotes a continuous-wave \ac{THz} source, L$_1$ is a converging lens, L$_2$ is 12~mm diameter focusing lens coupled with a detector, DOE$_{1}$ and DOE$_{2}$ are diffractive elements, T stands for a metallic target with periodic slits, and D denotes a detector. During imaging, the target T moves in the $xy$ plane. The silicon-based diffractive optical elements, conventional Fresnel zone plate (ZP), Fibonacci lens (F), Bessel axicon (B), and Airy zone plate (A), are placed in the positions indicated by DOE$_{1}$ and DOE$_{2}$ in various investigation scenarios. Schematics of the optical elements are depicted in the bottom left corner together with artistic depictions of the structured radiation they generate. On the right side, an opaque target with different period cut-out slits for an estimate of spatial resolution. In the bottom right corner, the sample is used for numerical simulations of the modulation transfer function.}
    \label{fig:Setups}
\end{figure}

The performance of diffractive optical elements (DOE) was investigated using the setup shown in Fig.~\ref{fig:Setups}. The investigation is based on an electronic multiplier chain-based emitter (\textit{Virginia Diodes, Inc.}) that provides \ac{CW} illumination of $0.5$~mW at 0.6~THz frequency ($\lambda=0.5$~mm). The converging lens was used to collimate the illumination, which was further engineered and collected by diffractive optical elements and then focused with 12~mm diameter high resistivity silicon lens coupled with the \ac{THz} sensor substrate based on an integrated log-spiral \ac{THz} antenna-coupled nanometric field effect transistor fabricated by a  90~nm CMOS foundry technology which exhibits a rather flat responsivity over the frequency range $0.1$--$1.5$~THz \cite{cibiraite2020}.  

The spatial resolution was experimentally estimated using a specially designed opaque target containing periodic slits with distances ranging from $0.3$~ mm to $2.5$~ mm. 

A high-resistivity silicon wafer of thickness $500$~\textmu m with a specific resistance greater than $10\,000~\Omega$~ cm and a refractive index of $n=3.46$ served as the core material for the fabrication of diffractive optical elements. The elements were designed for the wavelength $\lambda = 0.5$~mm corresponding to the frequency of $0.6$~THz and fabricated using the ultrashort pulsed laser ablation process \cite{Jokubauskis2018, Minkevicius2018}. 
For the latter process, a Pharos SP femtosecond pulsed laser (\textit{Light Conversion, Ltd.}) was employed; the ablation was done at wavelength $\lambda = 1030$~nm and beam size $9$~mm in the output at the intensity level of $\exp(-2)$. The optimal ablation parameters at the shortest pulse duration $\tau = 156$~fs were the average power $P = 5$~W at a repetition rate of $50$~kHz, thus producing $E = 100$~\textmu J energy per pulse. The beam was focused on the silicon surface with a fixed planoconvex lens $f = 100$~mm, while the substrates were installed using the ANT-XY95 precision stages (\textit{Aerotech, Inc.}). 
The pulse overlap density was $100$ pulses per millimeter, while the spot size was approximately $20$~\textmu m, thus removing the material height $0.9$~\textmu m in a single pass. 
This set of ablation parameters led to the best results in the fabrication of elements with a minimal heat oxidation zone, which allowed avoidance of silicon oxidation and maintenance of the surface roughness of the ablation at $< 2$~\textmu m.

\subsection{\label{sec:verif}Experimental verification of the diffractive optics elements}

\begin{figure}
    \centering
    \includegraphics[width=1\columnwidth]{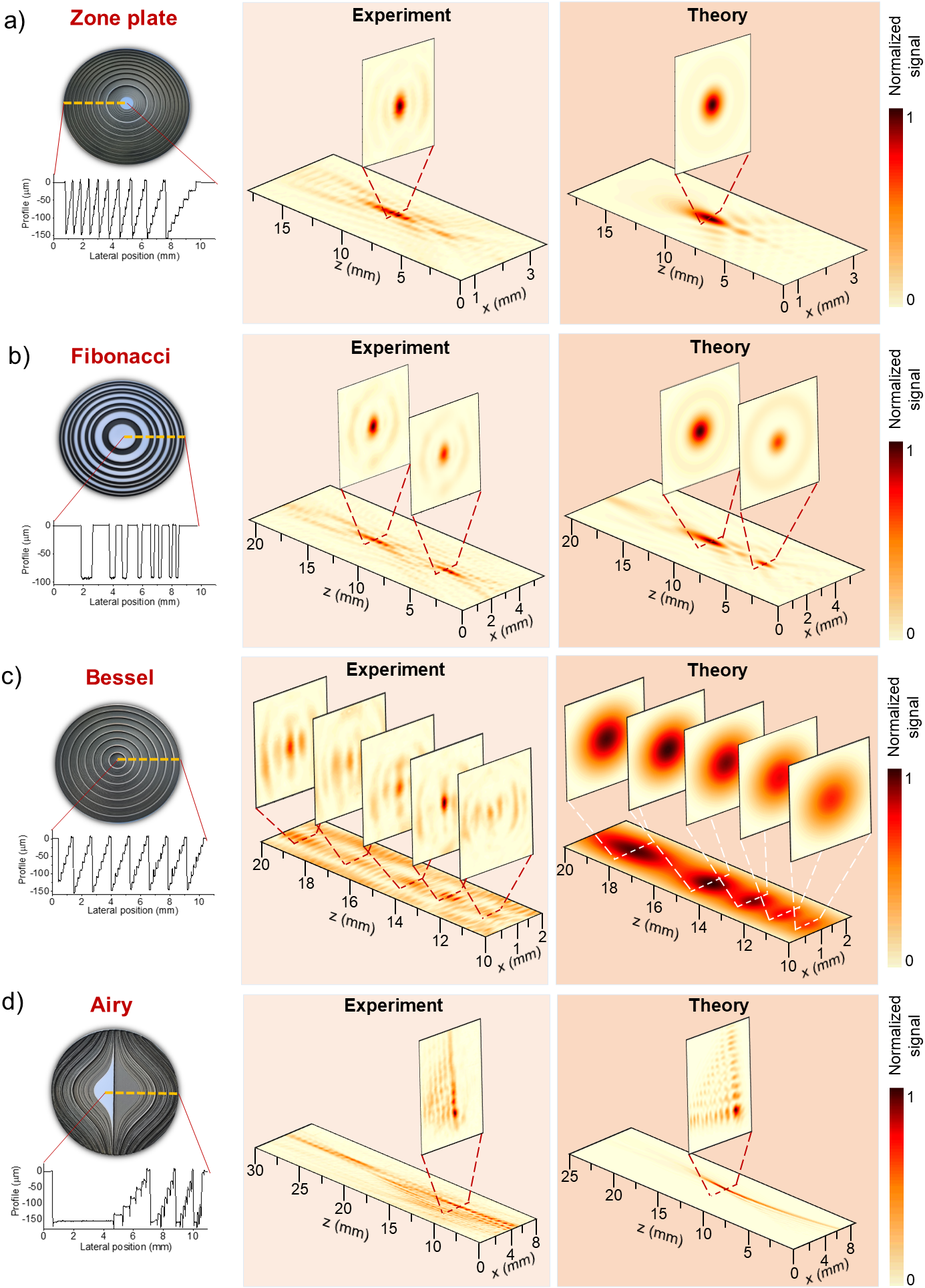}
    \caption{Fabricated diffractive optics elements (DOE) and their cross-sections together with experimentally generated light intensity profiles and profiles expected from the numerical calculations of (a) Fresnel zone plate with a clearly expressed focus in both plots; (b) Fibonacci lens displaying bifocal behaviour; (c) Bessel axicon producing 'needle'-shaped \ac{THz} beam and (d) Airy zone plate -- note a parabolic trajectory of propagation of the beam and the presence of reciprocal trajectories clearly pronounced in an experimental plot. A good correlation between modeling and experimental results is found. The intensity distributions are normalized to the maximal values of the signal.}
    \label{fig:Verifikacija}
\end{figure}

First, the performance of the elements produced was carefully examined. For this purpose, we chose to compare our numerical expectations with the actual performance of the diffractive photonic elements recorded in the experimental setup discussed above.

The diffractive silicon-based multilevel Fresnel \ac{ZP} Fig.~\ref{fig:Verifikacija}~(a) has been fabricated and studied experimentally using the \acs{THz}-\acs{CW} system described earlier, and the data obtained are compared with numerical estimates. The \ac{ZP} consists of concentric rings, the so-called Fresnel zones, which are spaced at such distances that light constructively interferes with the focal point located around $8.5$~mm from the element, as expected. An experimental investigation of the \ac{ZP} focusing performance was carried out by measuring the intensity distribution along the optical axis ($xz$) and in the focal plane ($xy$) perpendicular to the direction of \ac{THz} light propagation. A comparison of expected and recorded profiles is given in Fig.~\ref{fig:Verifikacija}~(a). The \ac{ZP} functions quite well -- in the focal plane, the \ac{FWHM} of the focused Gaussian beam is $0.27$~mm, while the \ac{FWHM} of the collimated beam amounts to $11.8$~mm; furthermore, the intensity of the focused beam increases $65$~times. The location of the focal point is consistent with expectations, despite some deviations, which are mainly due to the polarization profile of incident \ac{THz} light. 

Next, a Fibonacci lens is investigated. Its design is based on the aperiodic Fibonacci sequence principle, where each number is the sum of the two previous ones. In the Fibonacci lens, the ratio of nearby elements is closely related to the golden ratio. The main difference from the \ac{ZP} is that the Fibonacci lens is bifocal, where the spatial distance between the focuses follows the value of golden ratio \cite{golden-doi:10.1142/3595}. 
The Fibonacci beam is scanned by moving the detector in transverse planes ($xy$) and longitudinal planes ($xz$). The experimental results and their comparison with the theoretical model are presented in Fig.~\ref{fig:Verifikacija}~(b). As can be seen, the element forms two focal points in both cases, where the intensity in the second focus is $25 \%$ higher than in the first.
Furthermore, the \ac{FWHM} in the first focus was found to be $0.27$~mm. We guess this is caused by interference between the incident and reflected signals.

As a next step, we move on to the investigation of the axicon element dedicated to generating a Bessel beam. When the Gaussian beam passes through the axicon, it becomes engineered into a "needle". Thus, the focused \ac{THz} radiation is observed in a wide $\sim 10$~mm range of longitudinal distances, see Fig.~\ref{fig:Verifikacija}~(c). 
The experimentally obtained two-dimensional \ac{THz} Bessel beam profiles are given in Fig.~\ref{fig:Verifikacija}~(c). A total of five focal points are registered. The distance between focal points increases (from $\sim0.2$ to $1.5$~mm) with distance in the direction of propagation of the beam. Furthermore, as the beam expands, the \ac{FWHM} increases with each focal point, $1$~mm, $1.1$~mm, $1.9$~mm, $2.2$~mm, and $3.7$~mm, accordingly. The experimental results obtained are confirmed by numerical simulations, see Fig.~\ref{fig:Verifikacija}~(c). A precise look at the data reveals some deviations between numerical expectations and experimental observations. Our guess is that this is caused by interference between incident and reflected signals or by some not-perfect optical alignments in the experimental setup.

The property of the needle-shaped Bessel beam provides the advantage of not being so strict in the exact placement of optical elements and objects in the optical setup. Moreover, it deserves to be recalled that the Bessel beam is non-diffractive and self-healing. Therefore, during propagation, the beam does not spread out or diffract, and when encountering an obstacle, it will reappear after the obstacle (unless the obstacle is larger than the axicon) and continue to propagate in the initial direction. Theoretically, the amplitude of the Bessel beam is described by the Bessel function of the first kind.

Finally, we examine the performance of the Airy element; see Fig.~\ref{fig:Verifikacija}~(d). As depicted, the experimentally measured performance is in close line with what we expect from the numerical estimates: the Airy beam formed in the transverse and longitudinal planes displays a parabolic trajectory, which is an inherent feature of the Airy beams.
The beam in the transverse plane is scanned at the point of maximal longitudinal intensity and behaves as expected, although the edges are slightly distorted as a result of some deviations from the paraxial trajectory. 
It can be seen that in this case, Fig.~\ref{fig:Verifikacija}~(d), the Airy element was also combined with the zone plate. We note that similar measurements were reported with and without additional \ac{ZP} \cite{ivaskeviciute2022}. These results show that the \ac{ZP} behind the Airy zone plate slightly deflects the incident beam. Furthermore, the spatial resolution and quality of the scanned images were reported to be much better when an Airy zone plate is combined with a conventional \ac{ZP}.

Therefore, verification experiments and numerical examination allowed us to infer that the fabricated elements perform well and are well-suitable for precise investigation and benchmarking.

\section{\label{sec:results_and_discussion}Imaging with various single-pixel object inspection setups}

\subsection{\label{sec:imaging} Overview of paraxial vs nonparaxial and single-shot vs single-pixel imaging strategies}

An optical system commonly employs a specific type of target to evaluate its performance \cite{smith2008modern}. This target consists of a sequence of alternating light and dark bars, all with the same width. The system under examination captures multiple sets of patterns with varying distances between the bars. The finest set in which the line structure can still be perceived is regarded as the system's resolution limit, expressed as the number of lines per millimeter.

In the case of single-shot imaging \cite{duarte2008single} by an optical system, each infinitesimally narrow geometric line in the object appears as a blurred line in the image. The shape of this blurred line represents the line spread function. Traditionally, the image spread function smoothens the edges of the image, causing the image blur to have a more pronounced impact on increasingly finer patterns.
It becomes evident that if the contrast of the illumination in the image falls below the minimum detectable level of the system (e.g., the eye, film, or photodetector), the pattern can no longer be "resolved". 

\begin{figure}
    \centering
    \includegraphics[width=0.8\columnwidth]{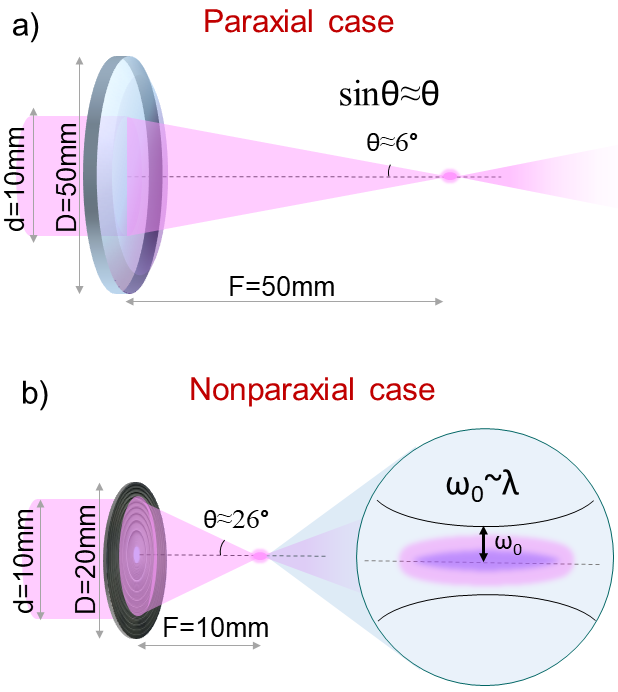}
    \caption{
    Schematics of paraxial case in panel (a), when the beam is focused with a conventional lens, and nonparaxial case (b), where the THz beam is focused with a diffractive optical element. The inset in panel (b) indicates the beam waist, which in this case is in orders of the \ac{THz} radiation wavelength.}
    \label{fig:Parax_nonparax}
\end{figure}

The backbone of modern optical engineering and imaging is geometrical optics. Geometrical optics, or ray optics, interprets light as rays, simplifying its behavior under specific conditions. Here, light travels in straight lines through a uniform medium, and it bends or splits when it encounters different mediums. Light follows curved paths only when the refractive index constantly changes and can be absorbed or reflected. However, geometrical optics does not consider effects coming from wave optics, such as diffraction and interference. This simplification is practical and highly accurate when the wavelength is much smaller than the size of the interacting structures, especially in imaging and optical aberration analysis. Yet another specific condition is the paraxiality of light rays - they propagate at small angles with respect to the optical axis, see Fig.~\ref{fig:Parax_nonparax}~(a), which in turn naturally leads to Gaussian optics \cite{smith2008modern}.

The situation becomes different when we move away from the optical frequencies to \ac{THz} because the characteristic wavelengths cannot be considered much smaller than the size of the interacting structures. The spatial frequencies of the \ac{THz} rays become relatively large and now do not propagate close to the optical axis. As a consequence, the numerical aperture $NA$ allows us to call this system a nonparaxial one, see Fig.~\ref{fig:Parax_nonparax}~(b). In particular, the scalar description of focal fields is no longer valid, and a variety of different analytical or semi-analytical models are used to describe focal fields \cite{orlov2010complex,orlov2014vectorial}. 

In our further considerations, we do estimate that our imaging setup is not paraxial anymore; however, we still assume that such electric field components as longitudinal or cross-polarized can be neglected.
We start by performing a numerical investigation of various imaging setups to evaluate their imaging properties. We perform a numerical imaging experiment using a sample containing a large group of four consecutive transparent and non-transparent stripes of ever-decreasing widths \cite{mundrys2023}. This sample, see the bottom inset in Fig.~\ref{fig:Setups}, enables us to determine the resolution of the object inspection setup with fine accuracy. 

Next, we introduce the concept of contrast to determine the resolution. To calculate the contrast at a given spatial frequency, the maximum intensity $I_{\max}$ is introduced, and $I_{\min}$ represents the minimum intensity.

\begin{equation}
\text { Contrast }=
\left[
\frac{I_{\max }-I_{\min }}{I_{\max }+I_{\min }}
\right] \times 100 \%.
\label{Eq:3}
\end{equation}

By representing the contrast in the image as a "modulation", we can create a plot showing how the transferred modulation varies with the number of lines per millimeter in the image. In conventional photo cameras, this plot is known as a \ac{MTF} indicating the ability of an optical system to transfer contrast at a particular resolution from the object to the image. 

The point where the modulation function line intersects a line representing the minimum detectable modulation level of the system sensor provides us with the limiting resolution of the system. For convenience, we set it in further experimentation as a constant at the level $20 \%$. In this way, we arrive at the description of the resolving power of the imaging setup.

\begin{figure}
    \centering
    \includegraphics[width=0.92\columnwidth]{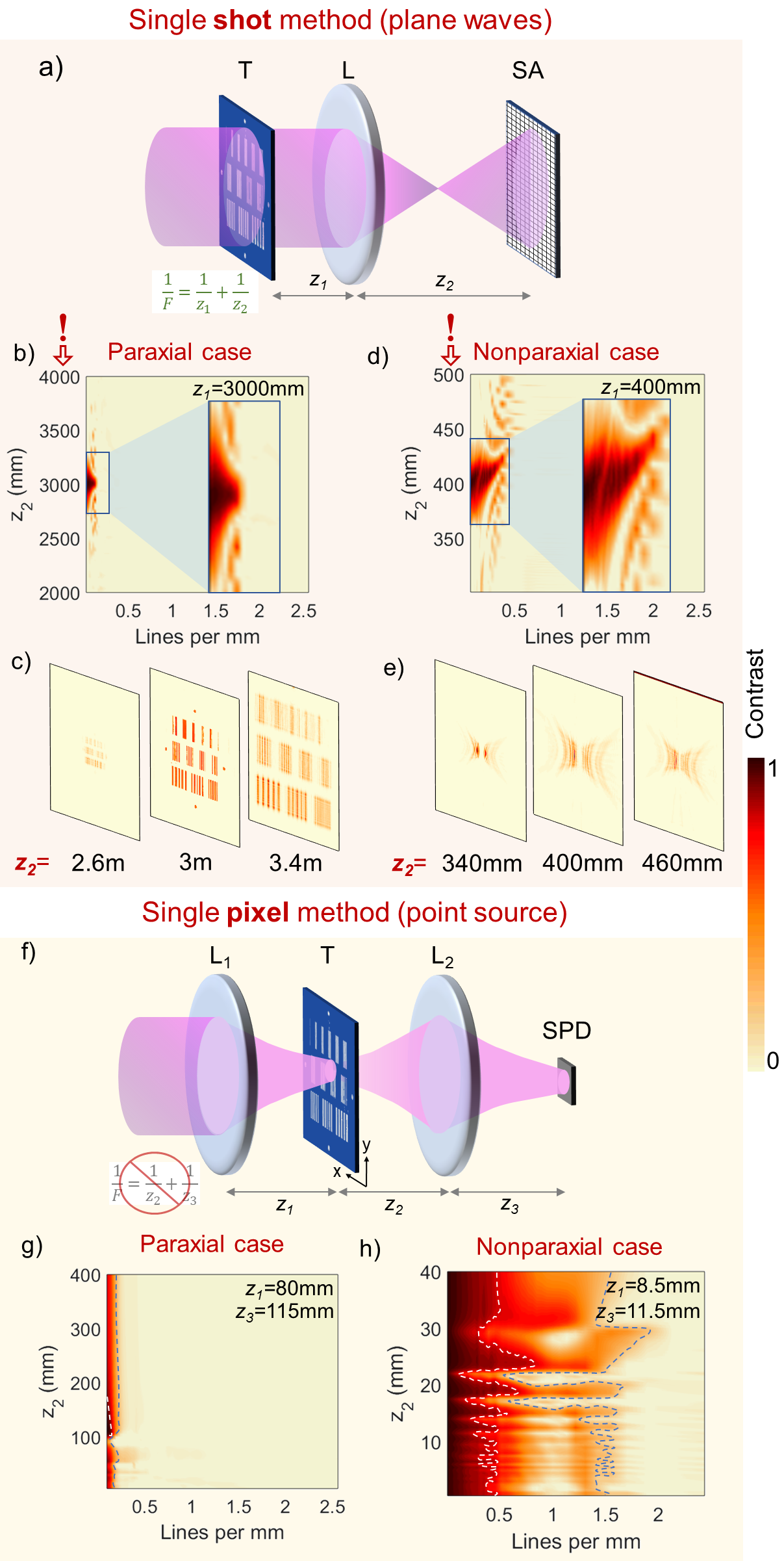}
    \caption{Schematic illustration of a single-shot (a) and single-pixel (f) imaging methods. The letter T indicates the target imaged, L, L$_1$ and L$_2$ denote lenses, SA labels a sensor array, and SPD marks a single-pixel detector. Numerically estimated modulation transfer function on the distance between the lens and sensor array for the single-shot imaging method and between the imaged target and the collecting lens for the single-pixel imaging shown in panels (b, d, g, h). Note the differences in the scales on the $z$ axis given in panels (b) and (d). Panels (b) and (g) demonstrate the modulation transfer map for the paraxial case, and (d) and (h) present the modulation transfer map for the nonparaxial case. Note also the distinction in scales on the $z$ axis given in panels (g) and (h). The white dashed outline represents a boundary where the contrast value exceeds $80 \%$ of the maximal value, the blue dashed outline marks the boundary of $20 \%$ of the contrast value in panel (h). Panels (c) and (e) present the single-shot target images in the paraxial and nonparaxial cases, respectively at $z_2$=2.6~m, 3.0~m, 3.4~m and 340~mm, 400~mm, 460~mm. 
    Animation to illustrate optical nonparaxiality is given as a separate file in the Supporting Information.} 
    \label{fig:single_shot_pixel}
\end{figure}

To reveal the peculiarities of compact \ac{THz} imaging systems, we explore two different recording geometries in the setups, single-shot (collimated beam) and single-pixel (focused beam) in both paraxial and nonparaxial optics cases. Their schematic illustration is given in Fig.~\ref{fig:single_shot_pixel}. Single-shot imaging corresponds to the way the image is performed with Gaussian optics \cite{iizuka2013engineering,smith2008modern}. A target is illuminated with homogeneous illumination; the signal that is reflected or transmitted through the sample and directed to the lens forms an image of the object on the sensor array of single-pixels, see Fig.~\ref{fig:single_shot_pixel}~(a).

In the paraxial case, the image of the object on the sensor array is undistorted, though it can be magnified or minimized with respect to the size of the object. However, nonparaxial imaging will largely distort the outer parts of the image. This effect is studied in the theory of aberrations \cite{welford2017aberrations}. To exemplify this difference, we performed a numerical experiment in which we performed a raster scan of the sample (see the bottom inset in the last image in Fig.~\ref{fig:Setups}) and determined whether the central group of black-white strips is resolved on the on-axis. The results for two cases of a paraxial lens ($f=1500$~mm) and the nonparaxial one ($f=200$~mm) are given in Fig.~\ref{fig:single_shot_pixel}~(b,c). 

We see that according to the formula $z_1^{-1}+z_2^{-1}=f^{-1}$, for a particular position of $z_1=3000$~mm, the modulation transfer function reveals a single distinct position $z_2$ equal to $z_1$, as expected. Furthermore, from this contrast map we can estimate the \ac{DOF} of the paraxial imaging system, see Fig.~\ref{fig:single_shot_pixel}~(b) -- it is the highest contrast area distributed along the $z_2$ axis. 

The \ac{DOF} in an imaging system refers to the ability to maintain a desired level of image quality, specifically in terms of spatial frequency at a specified contrast, without requiring refocusing when the position of the object is adjusted closer to or further from the optimal. When an object is positioned either closer to or farther from the designated focus distance, the object's sharpness deteriorates, leading to a decrease in both resolution and contrast. Thus, depth of field is a term that describes how much of an image is "in focus", from the nearest to the farthest point. Therefore, the \ac{DOF} is meaningful only when defined in conjunction with a specified resolution and contrast.  The \ac{DOF} is not a fixed value, but rather a range of acceptable sharpness that varies depending on the viewing conditions and the desired image quality. To measure and compare the \acp{DOF} of different diffractive elements, we need to define some criteria for what constitutes acceptable sharpness. 

One way to do this is to use a target with known details or features of different sizes and heights and observe how they appear in the image as the distance from the collecting element changes. To assess and evaluate an imaging system's \ac{DOF} we use specific targets for direct measurement and numerical benchmarking. Quantifying \ac{DOF} is challenging without specifying the size of the object details or the spatial frequency within the image space. Generally, smaller details of the object $dx$ require higher spatial frequencies $\sim 1/dx$, which, in turn, results in a narrower \ac{DOF} than the imaging setup can achieve. The smaller the detail or feature, the higher the spatial frequency (the number of lines per unit distance) it represents, and the more sensitive it is to blurring. Therefore, we can use spatial frequency as a proxy for detail size and contrast as a measure of how well detail is resolved in the image. To visualize how a lens performs across a given range of depths at a particular detail size, a modulation transfer map is used. It can help us visualize how a lens performs over a certain depth range at a specific detail size. A cross-section of the MTF map has two main parts: a peak and a tail. The peak represents the plane of best focus, where the contrast is the highest. The tail represents the regions where the contrast drops below a certain threshold, indicating that the image is no longer acceptably sharp. The distance between the two tails is the \ac{DOF} of that spatial frequency. For simplicity, we set the contrast requirement for the definition of a \ac{DOF} as $80~\%$.

We apply these considerations to the MTF in Fig.~\ref{fig:single_shot_pixel}~(b) and the \ac{DOF} is estimated in this case to be rather modest. The shape of the high-contrast region around the "in focus" position has a distinct triangular shape, with some side lobes before and after that particular position on the $z$ axis.

For further discussion, we define the spatial frequencies $\xi = k/(2 \pi)$. According to the wavelength $\lambda = 0.5$~mm, the spatial frequency $\xi = 2$~mm$^{-1}$ means that we resolve objects comparable to the wavelength. We give the following numbers as guidelines to decide whether the spatial frequencies are low, intermediate, or high. We call spatial frequencies $\xi \approx 0.1~\text{mm}^{-1}$ low, as they correspond to objects around $20 \lambda$. Intermediate (or moderate) spatial frequencies are defined in the range $\xi \in (0.5, 1.33)~\text{mm}^{-1}$. They correspond to the resolvable details from $4 \lambda$ to $1.5 \lambda$. Spatial frequencies greater than $\xi= 1.33$ we will call high spatial frequencies.

When we move on to the example of the nonparaxial lens, see Fig.~\ref{fig:single_shot_pixel}~(d), we observe two distinct effects on the modulation transfer map: a) the resolving power increased due to the smaller focal length and b) the particular zone of good contrast became distorted. These effects will be amplified more as the focal lengths decrease. Note that the \ac{DOF}, in this case, is larger; however, in microscopy and similar applications, this fact is not useful because of the aim of best-resolving power in a single plane.

With this in mind, we now proceed to the single-pixel imaging scenario Fig.~\ref{fig:single_shot_pixel}~(f). We investigate here two scenarios: one with a paraxial lens and larger distances, see Fig.~\ref{fig:single_shot_pixel}~(g) and one with smaller distances and Fresnel zone plates (we will discuss it in detail in the next Section), see Fig.~\ref{fig:single_shot_pixel}~(h).

We immediately notice that in single-pixel paraxial imaging, there is no distinct field of view, see  Fig.~\ref{fig:single_shot_pixel}~(g). The axial positions $z_2$ produce a relatively large area with some improvements at a particular distance of around $z_2=100$~mm. Most notably, the resolution power is lower than that of a single shot system, compared to Fig.~\ref{fig:single_shot_pixel}~(b).

The depiction of the modulation transfer function from the next Section explains in the first place why this research is so valuable -- we observe a large number of areas on the axis, where the resolving power of the nonparaxial \ac{THz} single-pixel imaging system is high, see Fig.~\ref{fig:single_shot_pixel}~(h). The classical formula $z_1^{-1}+z_2^{-1}=f^{-1}$, which comes from single-shot conventional imaging, appears to be no longer valid. In the following, we will discuss whether rational design rules applicable to single-pixel \ac{THz} imaging can be established.

Lastly, before proceeding to the next Sections, it is worth noting that the limiting resolution alone does not completely characterize the system's performance. Two modulation plots can have the same limiting resolution but exhibit significantly different performances. The plot showing higher modulation at lower frequencies is clearly superior, as it produces sharper and more contrasting images \cite{nasse2008read}. Unfortunately, when comparing two systems, the choice is often not as straightforward. For instance, one system may have a high limiting resolution, while the other demonstrates high contrast at low target frequencies. 

In such cases, the decision must consider the relative importance of contrast versus resolution in relation to the intended function of the imaging system. Therefore, with this in mind, we proceed to evaluate individual single-pixel imaging setups.

\subsection{\label{sec:gaussian_modes}Conventional imaging setup using lenses}

The most common imaging setup, as a rule, involves one or two lenses and is a must-have for single-shot imaging when the incident beam is collimated on the target \cite{iizuka2013engineering,born2013principles}. A schematic illustration is given in Fig.~\ref{fig:single_shot_pixel}~(a). The distances $z_1$, $z_2$ between the object and the lens and between the image and the lens, respectively, are related by formulae $z_1^{-1}+z_2^{-1}=f^{-1}$, where $f$ denotes the focal length. This equation governs the best contrast, resolution, and image brightness. However, as we did note in the previous discussion, it is valid for conventional, i.e.\ paraxial imaging systems \cite{iizuka2013engineering,smith2008modern}, and begins to fail because lenses and objects involve nonparaxial diffraction theory, as evidenced by the plots of Fig.~\ref{fig:single_shot_pixel}~(b,d). This can be corrected by the introduction of vectorial diffraction theory for high-numerical aperture systems \cite{braat2008assessment}.

In this study, we focus on the analysis of the aforementioned relation for various nonparaxial imaging scenarios as the fabricated silicon zone plates are nonparaxial. This issue is of particular importance in the future development of reduced-size single-pixel \ac{THz} imaging systems, and, especially, when optical elements can be designed to be integrated with compact \ac{THz} sensors on a single chip.

\begin{figure}
\centering
 \includegraphics[width=0.95\columnwidth]{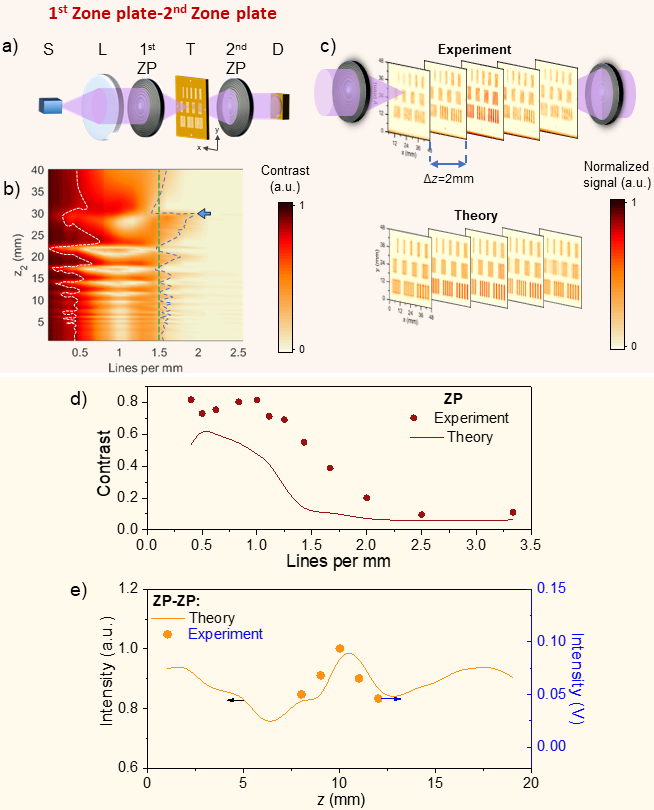}
 \caption{Conventional \ac{THz} imaging of a target using two diffractive \acp{ZP}. Panel (a) -- experimental setup, where S denotes a \ac{THz} source, L stands for a converging lens, 1$^{\text{st}}$~ZP and 2$^{\text{nd}}$~ZP marks the positions of the two identical \ac{ZP}, T denotes a target, and D is a detector; panel (b) -- numerically estimated modulation transfer map on the distance between the target and 2$^{\text{nd}}$~ZP. The white dashed outline represents the boundary where the contrast value is greater than $80 \%$ and is used to estimate the \ac{DOF} of the setup. Note triangular-shaped areas with good contrast. For higher spatial frequencies up to $1.5$~lin/mm, indicated as a guide for the eye by a straight dashed green line; a dashed blue outline marks the boundary of $20~\%$ of the contrast value, the blue arrow labels the area of the highest resolution -- it reaches $2$~lin/mm; panel (c) shows experimentally recorded (top image) and numerically estimated (bottom image) \ac{THz} images of the target, using two \acp{ZP}. Distributions are normalized to the maximum value of the signal; panel (d) depicts experimentally and numerically obtained modulation transfer function; panel (e) -- dependencies of the image brightness on the position of the target for the fixed distance between the zone plates, 1$^{\text{st}}$~ZP and 2$^{\text{nd}}$~ZP.}
\label{fig:ZP}
\end{figure}

The principal optical scheme of the experiments performed is shown in Fig.~\ref{fig:ZP}~(a), where the incident \ac{THz} beam passes through the converging high-density polyethylene (HDPE) lens L, with focus $F=12$~cm, and illuminates the diffractive zone plate (ZP) that forms a Gaussian beam, which finally reaches the target (T). For imaging purposes and its quality estimates, a $5 \times 5$~cm metal plate with cut bars of a different period ($0.3$ -- $2.5$~mm) has served in the experiments, see Fig.~\ref{fig:Setups}. The period of bars that can be distinguished allows us to experimentally determine the spatial resolution. Then, the beam passing through the target reaches the second zone plate element (2ZP), which focuses it on the detector (D). 
All experimental two-dimensional profiles were obtained by raster scanning the target at a speed of $10$~mm/s, while the scattered \ac{THz} light was collected by the second \ac{ZP} to direct it to the detector. 


In our numerical simulations, we use two objects: the first is a sample for the modeling of the \ac{MTF} and determination of the resolution (see the bottom inset in Fig.~\ref{fig:Setups}), and the second, the aforementioned target with different periods of cut-out slits for comparison with the experiment. Within the numerical experiment, both samples are kept at a distance of $8.5$~mm from the first illuminating lens. The resulting contrast of the image is examined in the modulation transfer function \cite{smith2008modern}, which is shown on a modulation transfer map. The results of this numerical experiment are presented in Fig.~\ref{fig:ZP}~(b). First, it is worth noting the presence of several regions in the $z$ scale that are suitable for placing the light-collecting lens with relatively high contrast. A distinct difference between the single-shot and single-pixel imaging versions is clearly observed when comparing Fig.~\ref{fig:single_shot_pixel}
and Fig.~\ref{fig:ZP}. 

Analogously to single-shot imaging, we can speak about the \ac{DOF} of the imaging system in nonparaxial single-pixel imaging. The \ac{DOF} is defined as the distance between the closest and furthest elements in a scene that appear to be 'acceptably sharp' in the image. These distances can be extracted from the \ac{MTF} maps, similar to that in Fig.~\ref{fig:ZP}~(b). As we did notice, there are a number of on-axis positions with a high resolution of lines per millimeter in the numerical sample. The \ac{DOF} around these particular positions can be seen as triangular areas with an equidistant level of contrast in Fig.~\ref{fig:ZP}~(b), see, for instance, areas above $10$~mm, $12$ -- $14$~mm, around $15$~mm, around $25$~mm and $32$ -- $40$~mm. Therefore, when the object is nearer ($z_2$ is smaller), the \ac{DOF} is smaller too, and when the imaged sample is further away, the \ac{DOF} is accordingly larger. These areas denote positions where the lower and intermediate spatial frequencies are resolved with high contrast.


Whereas in the single-shot imaging, there is a single possible good position, where to place the second \ac{ZP}; for the single-pixel imaging, one can clearly resolve several positions with good contrast for high spatial frequencies up to $1.5$ lines per millimeter (lin/mm), marked as a guide for the eye by a dashed green line. However, if we aim for resolutions better than that, there are few particular regions of distances with good contrast, most notable being located around $z = 24$~mm and $z=30$~mm with expressed resolution reaching $2~$lin/mm, see the contour line in Fig.~\ref{fig:ZP}(b). As we know the distance between the detector and the second \ac{ZP}, we can conclude that \textbf{these distances do not obey the classical single-shot imaging law.} For the first case, $z_1^{-1}+z_2^{-1} = 0.129~\text{mm}^{-1}$ gives the effective $f = 7.7$~mm, and for the second case $z_1^{-1}+z_2^{-1} = 0.121~\text{mm}^{-1}$ resulting in $f = 8.26$~mm, which is less than the actual focal point distance of $8.5$~mm and the designed paraxial focal length of $10$~mm.

In the next experimentation, we experimentally measure the \ac{DOF} of the imaging setup. We fix the distance between two \acp{ZP} and move the object, looking for the brightest spot. The experimental and numerical results of the investigated \ac{ZP} performance are shown in Fig.~\ref{fig:ZP}~(c). The colored scale is normalized to the maximum signal. To experimentally examine the imaging performance using \acp{ZP}, a non-transparent sample is raster scanned (Fig.~\ref{fig:ZP}~(c)). To determine \ac{DOF}, the sample is also moved between both optical elements with a step of $1$~mm. As one can see, the experimental data are well supported by the numerical calculations.

\begin{figure*}
    \centering
    \includegraphics[width=0.95\textwidth]{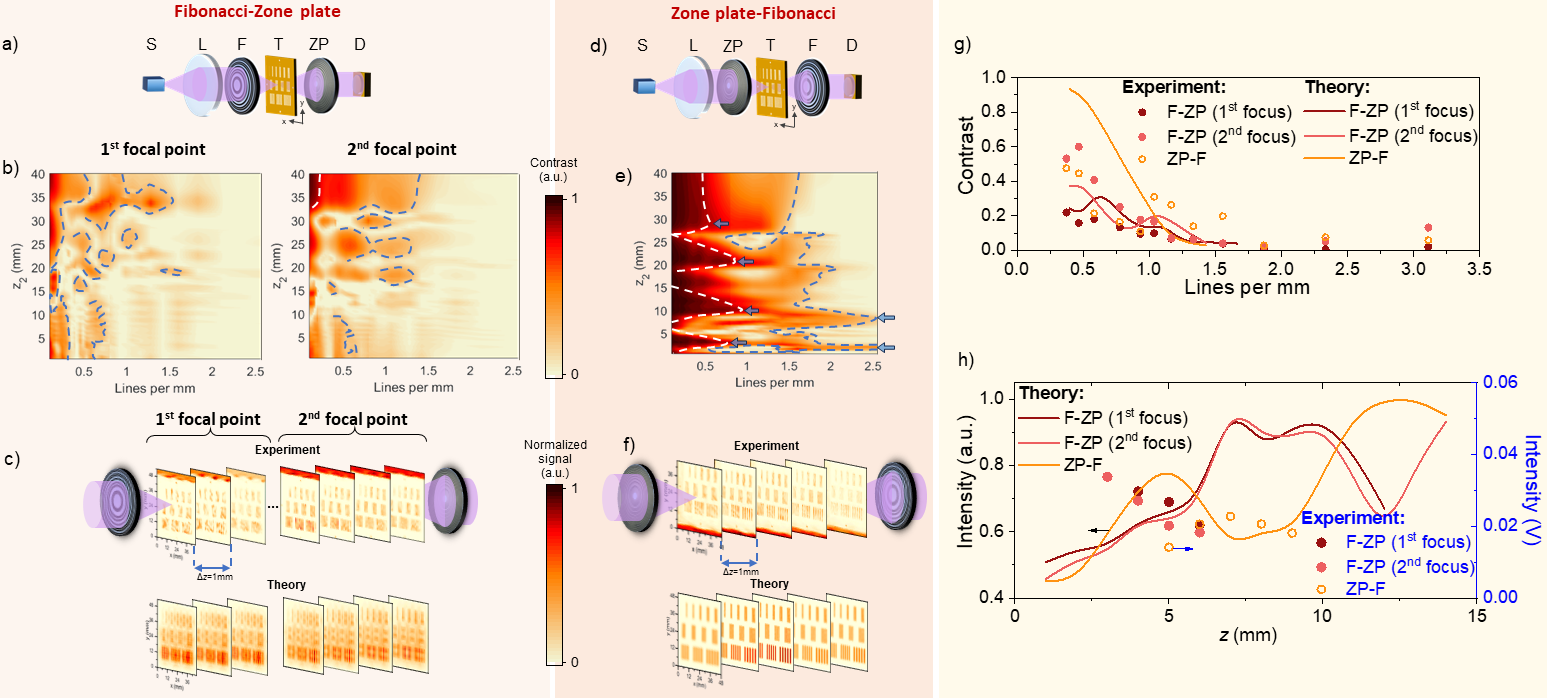}
    \caption{Imaging of the target using the Fibonacci lens and a diffractive zone plate in two different setups: (a) when the Fibonacci lens serves as an illuminating and ZP as a collecting element, (d) reverse combination of the optical components in the light illumination and collection scheme. In both setups: S is a \ac{THz} source, L denotes the converging lens, F is the Fibonacci lens, ZP stands for zone plate, T is a target and D is a detector. Panels (b) and (e) -- numerically estimated modulation transfer map on the distance between the target and the second optical element. Note that at the first focal point, $z_2 \sim 33$~mm the spatial frequencies correspond to up $0.8$~lin/mm. In panel (e) the white dashed outline represents the boundary where the contrast value is greater than $80 \%$ -- note clearly resolved triangle-shaped areas; the dashed blue outline marks the boundary of $20 \%$ of the contrast value, the blue arrow labels the area of the highest resolution -- its value was found to be 2.5~lin/mm; panels (c) and (f) -- images obtained experimentally (top picture) and numerically estimated (bottom picture) \ac{THz} imaging of the target. Distributions are normalized to the maximum value of the signal. Panel (g) presents experimentally and numerically obtained modulation transfer functions; panel (h) shows the dependence of intensity on the distance between the first optical element and the imaged target.}
    \label{fig:Fibonacci}
\end{figure*}

The position where the image of the object is the most bright was found to be when the object is in the middle between the first and second ZPs. For this particular distance, we plot both the experimental and numerically determined contrast in the form of \ac{MTF}, see Fig.~\ref{fig:ZP}~(d). The spatial resolution estimated experimentally is $1 \lambda$. Furthermore, a high \ac{SNR} of $3136$ was recorded. Numerical expectations are generally similar; however, the numerically predicted spatial resolution is smaller, probably because of slightly different distances between experiments and those used in theory.

Another important indicator of the quality of the setup is the maximum brightness of the image recorded by a detector as the intensity. The proper term for image intensity is image irradiance \cite{jain1995machine}, but terms such as intensity and brightness are so common that they have been used as synonyms for image irradiance throughout this text. We found that it depends on the distance between the object and the second \ac{ZP}, see Fig.~\ref{fig:ZP}~(e). The experimental results were supported by numerical simulations. This measurement is particularly important because during experimentation the position of the sample to be placed is found while searching for the optimal (i.e.\ the strongest) signal. However, \textbf{the optimal modulation transfer condition appears to be decoupled from the condition for the best image irradiance}, compare Fig.~\ref{fig:ZP} (b) and (e). We recently reported on this finding using mean square error image inspection metrics between the image and the object in numerical experiments \cite{mundrys2023} and now confirm it experimentally. 

Thus, we observed \textbf{differences between the single-shot and single-pixel imaging principles}. 
This observation has motivated us to investigate further numerous single-pixel imaging scenarios.

\subsection{\label{sec:bifocal_beam_engineering}Single-pixel bifocal imaging using a single Fibonacci lens}

As a next step, we are curious about the performance of the Fibonacci lens, as it creates two Gaussian-like focal spots. Therefore, it is intriguing to ask whether imaging is possible in a setup using the bifocal lens for the illumination of the target for the light collection and image formation.

In the case of single-shot imaging, the performance of the bifocal element is expected to be worse than that of the \ac{ZP} element, as the element collects light from two focal points onto a sensing element \cite{ferrando2014imaging}. However, the question is not trivial 
and is open to a single-pixel raster scan scheme. 

For this investigation, we choose two different scenarios; see Fig.~\ref{fig:Fibonacci} -- the first is when the bifocal element is used to create structured \ac{THz} illumination on the sample, and the second is when it is used to collect the light coming from the sample to the detector. In the first case, see Fig.~ \ref{fig:Fibonacci}~(a), we placed the sample in two different focal planes and still collected light to direct it to the detector with the lens; see Fig.~\ref{fig:Fibonacci}~(b). Our naive expectation is that the illumination will have no effect on the contrast determined in the numerical experimentation, as we still move the imaging lens away from the object. To our surprise, changes in the structure of the illumination have a distinct effect even when the sample was placed in different spots of the bifocal element; see Fig.~\ref{fig:Fibonacci}(b). For target inspection, the first focal point of the Fibonacci lens results in a decreased spatial resolution. Pay attention to the low spatial frequencies, we observe some distances between the object and the imaging lens for which the target could be resolved. However, the modulation transfer map is strongly inhomogeneous, and, moving to the intermediate spatial frequencies, the contrast at those particular positions drops and increases afterward, see dashed line in Fig.~\ref{fig:Fibonacci}(b). Of course, some regions of the \ac{MTF} with the improved contrast are present for regions of the sample with structures containing 1 to 1.5~lin/mm. Unfortunately, smaller structures do not resolve as well, notice regions with contrast below our limit of $20 \%$. As \ac{THz} light is collected using \ac{ZP}, we estimate $z_1^{-1}+z_2^{-1} = 0.116~\text{mm}^{-1}$, so it gives the effective $f = 8.6$~mm, which is slightly larger than expected. 

The situation shows improvements as we place the sample at the second focal point; see Fig.~\ref{fig:Fibonacci}~(b), right panel. Here, we observe some similarities with the numerically detected imaging tendency from the case discussed in the previous Section. There are several different positions for high-resolution imaging, which exceeds the value $20 \%$ of contrast; see, for instance, positions at $z_2 \sim 15$~mm when the spatial frequency of 0.4~lin/mm is resolved, $z_2 \sim 18$~mm and $z_2 \sim 25$~mm which corresponds to the spatial frequency of $1.4$~lin/mm and $z_2 \sim 38$~mm that reaches the spatial frequency $1.5$~ lin / mm. The MTF map is enhanced, although the best-case scenario shows a decrease in resolution, compared to Fig.~\ref{fig:ZP}. The best position of the \ac{ZP} imaging is $z_2=26$~mm, from which we estimate $z_1^{-1}+z_2^{-1} = 0.126~\text{mm}^{-1}$, and the effective focal length is $f = 7.92$~mm, which is less than expected. 

The modulation transfer map in Fig.~\ref{fig:Fibonacci}~(b) also provides some crucial information on the \ac{DOF} of the imaging system. As we can conclude from the inspection of this figure, the \ac{DOF} varies. It can stretch from a few millimeters ($3$--$4$~mm) for focal points around $z_2=17.5$~mm (the first focal point) and $z_2 = 15$~mm (the second focal point) to tens of millimeters at distances around $z_2 = 26.5$~mm (the first focal point) and $z_2 = 40$~mm (the second focal point).

The focusing performance of the Fibonacci lens was investigated both experimentally and numerically; see Fig.~\ref{fig:Fibonacci}~(c). In this experimentation, we fix the distances between the illumination and light-collecting elements, but vary the position of the target. The same target is used to examine the imaging ability using Fibonacci illumination (see Fig.~\ref{fig:Fibonacci}). As in the previous case, the target is scanned by a raster moving it in the transverse plane ($xy$) and repeating the process by moving the sample in the longitudinal direction ($z$) with a step of $1$~ mm. In this way, we are able to estimate the contrast, resolution, and DOF experimentally for both focal points. As can be seen in Fig.~\ref{fig:Fibonacci}~(c), illumination at the second focal point results in a more intense signal and a better overall target imaging quality is also observed at the second focal point, supporting the claims of the numerical simulation. The \ac{SNR} at the first focal point was found to be $595$; meanwhile, the \ac{SNR} is much higher, around $1000$, and was estimated at the second focal point. Compared to the scenario discussed previously with two \acp{ZP}, it is still $3$ times smaller. The experimental results agree well with the numerical simulations.

Slightly puzzled by this behavior, we change the positions of the lens and the bifocal element, Fig.~\ref{fig:Fibonacci}~(d). Thus, the \ac{THz} structured light illumination is now created using a common zone plate, but the light-collecting and imaging-performing element is a Fibonacci lens. Here, we see an improvement in the situation compared to the nonswitched case. At least four positions of the imaging element are present, where the contrast increases for regions of the sample with a high number of lines per millimeter; see Fig.~\ref{fig:Fibonacci}~(e). Surprisingly, the Fibonacci element gives sharp images even if placed near the sample at a distance $z \approx 3$~mm. The second good position to place a bifocal element is at a distance around $z \approx 9-10$~mm. The next distance, where the element performs high-resolution imaging, is $z \approx 20$~mm. A slight drop in resolution is observed at $z \approx 28$~mm, although the resolution drops slightly compared to the distances mentioned above; compare Fig.~\ref{fig:Fibonacci}~(e) and Fig.~\ref{fig:ZP}~(b).

In quite a fascinating fashion, the \ac{DOF} is different now, too. The importance of the illuminating object is clearly revealed while comparing Fig.~\ref{fig:Fibonacci}~(b) and Fig.~\ref{fig:Fibonacci}~(e). First of all, we observe fewer axial positions where spatial resolution is good. The depth of field of those particular positions is also increased, compared to the previous case. The shape of those regions is a distorted triangle with a better resolution of objects that are further away from the optimal spots. This contrasts with the more or less symmetrical shape of the \ac{DOF} zone for single-shot imaging. In particular, we observe the presence of four different axial zones that we can call focal points due to the good resolving power, see Fig.~\ref{fig:Fibonacci}~(e). These particular locations are at $z_2 \sim 2.5$~mm, $z_2 \sim 12.5$~mm, $z_2 \sim 22.5$~mm, and $z_2 \sim 27.5$~mm. Note that the shape of the \ac{DOF} region is distorted: distances, where fine details could be resolved, are generally closer than distances when poor resolution is predicted.
Low spatial frequencies are optimally resolved in all focal spots, with intermediate frequencies detected at some locations. Note that even high spatial frequencies up to $2.5$~lin/mm can be detected within this imaging setup.

These three examples demonstrate how restricted the evaluation of system performance is based solely on the limiting resolution. The MTF plots have points with identical limiting resolutions and exhibit significantly different performances. The plot that displays higher modulation at lower spatial frequencies clearly outperforms the others, as it will produce sharper and more contrast-rich images. Unfortunately, the decision-making process and the selection of configuration become less straightforward when faced with choosing between numerous systems. For instance, one system displays high limiting resolution, while others may demonstrate high contrast at low target frequencies. In such cases, the decision must be guided by determining the relative importance of contrast versus resolution in relation to the intended function of the system.

To verify these findings, we performed an experimental verification for the fixed distance between the illumination and imaging elements while the position of the sample changed; see Fig.~\ref{fig:Fibonacci}~(f). This time, the collimating direction is switched with the collecting direction, meaning that the Fibonacci lens was placed behind the target. In this experiment, the Gaussian beam passes through the imaged target, passes through the Fibonacci lens, and reaches the detector. The \ac{SNR} of the image at the focal point is $344$. Compared to the previous combination of lenses, it is $43 \%$ less than at the first focal point and by $66 \%$ less than an \ac{SNR} at the second focal point.

Experimentally determined resolution, contrast, and depth of field are supported by numerical experimentation using the same parameters as in the experiment performed. Therefore, in general, a good agreement can be indicated between the experimental result and the expectations from the numerical investigation.

We select the cases with the best resolution and plot the contrast dependence of the regions in the sample with different numbers of lines per millimeter in the substructure; see Fig.~\ref{fig:Fibonacci}~(g). Some slight deviations are present, but overall the tendencies are similar.

Once again, resolution is not the single parameter that qualifies the performance of the imaging system. Irradiance is also important for this task; therefore, we present both experimental and numerical results in Fig.~\ref{fig:Fibonacci}~(h). For the first scenario, when the bifocal element is used to structure the \ac{THz} light and illuminate the sample, we observe a single brightness spike when the lens is at $z=7-10$~mm, which more or less corresponds to the expected focal distance of the \ac{ZP} collecting light. When the bifocal element is used to perform an imaging task, two distinct regions are present, from which the \ac{THz} light is optimally collected onto the detector. Therefore, in these cases, the optimal modulation transfer condition is also independent of the condition for the best irradiance; compare Figs.~\ref{fig:Fibonacci}~(b,e) and (h).

Most importantly, we observe the distinct difference in object inspection scenarios when changes are made not only in the structure of the illumination but also in the imaging element.

\subsection{\label{sec:bessel_zone_plate}Single-pixel imaging using  axicon}

Lenses are not the only optical elements that can focus light. The axicons do not create a single focal spot, but rather a line of focused light. For single-shot imaging, this implies that an image will be created from objects located on that line (known as a Bessel zone), so the distances will not be distinct \cite{tanaka2000comparison, zhai2009extended}. Usually, such images have to be digitally post-processed to resolve the investigated objects \cite{zhai2009extended}.

\begin{figure}
    \centering
    \includegraphics[width=1\columnwidth]{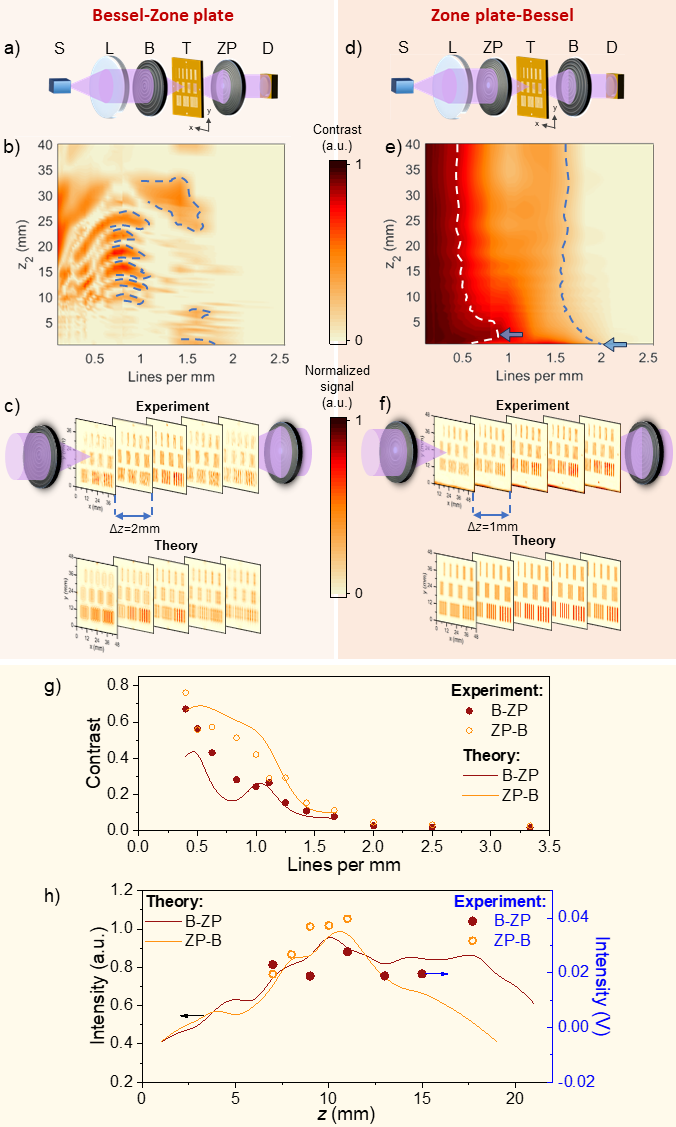}
    \caption{Target imaging using axicon and the diffractive ZP in two different setups: (a) when the axicon serves as an illuminating and ZP plate as a collecting element, (d) a reverse combination -- ZP serves as illuminating element and the axicon - as collecting. In both setups: S is a \ac{THz} source, L denotes the converging lens, B is an axicon, ZP stands for zone plate, T is the imaged target, and D is a detector. Panels (b) and (e) -- numerically estimated modulation transfer function on the distance between the target and the second optical element for both setups. In panel (b), a green dashed outline denotes an area with improved resolution. In panel (e), a white dashed line represents the boundary where the contrast value is greater than $80 \%$ of the maximal value; a dashed blue outline marks the boundary of $20 \%$ of the contrast value, and a blue arrow labels the area of the highest resolution reaching $2$~lin/mm. Note that the range is rather extended -- up to 40~mm range of high resolution of $1.7$~lin/mm. Panels (c) and (f) -- objective images obtained experimentally (top picture) and numerically estimated (bottom picture) target \ac{THz} imaging. Distributions are normalized to the maximum value of the signal. Panel (g) shows the numerical and experimental modulation transfer function; panel (h) -- dependence of intensity on the distance between the first optical element and the imaged target.}
    \label{fig:Bessel}
\end{figure}

Single-pixel imaging is a rather different type of fellow -- it has already been reported that axicons can improve resolution \cite{minkevicius2019bessel}, especially in conjunction with the mathematical deconvolution procedure. Here, we introduce in the imaging system one axicon, which is used together with a lens, for illumination or imaging purposes; see Fig.~\ref{fig:Bessel}.

Our interest here is to start by replacing the illuminating zone plate with an axicon. This will automatically structure the nonparaxial illumination; however, since the imaging element is still a zone plate, we wonder what imaging tendencies and principles will be valid here. The results of the numerical experiment are given in Figure~\ref{fig:Bessel}~(b). First, we note regions where the contrast is drastically decreased. However, regions of high contrast are present in the modulation transfer map, which means that Bessel illumination can improve the quality of the image. Especially interesting is the fact that the contrast for finer structures can be better than for larger structures; see Fig.~\ref{fig:Bessel}~(b). The most optimal position for the light collecting element here is around $z = 25$~mm; thus, we estimate $z_1^{-1}+z_2^{-1} = 0.127~\text{mm}^{-1}$, so it gives the effective $f = 7.8$~mm, which is less than expected (we recall that \ac{ZP} focuses the light at the distance $8.5$~mm) and is more notably different from the cases discussed previously. Finally, we note that the good contrast areas are not uniform with respect to the position of the light-collecting zone plate.

The illumination of the axicon is different from other types of illumination as it is created with a distinct spatial frequency $\xi = 2 \sin \beta~\text{mm}^{-1}$ or $\xi = 0.78~\text{mm}^{-1}$. This fact finds its manifestation in Fig.~\ref{fig:Bessel}~(b). Note the region of on-axis locations, where this particular frequency is resolved better than the lower spatial details in the sample. It is also noteworthy that there is also an area of locations in the modulation transfer map, where the higher spatial details are reasonably well resolved; see Fig.~\ref{fig:Bessel}~(b).

The modulation transfer dependence enables us to analyze the \ac{DOF} of this particular setup. Most notable is the fact that the definition of \ac{DOF} is troublesome for this particular case, as the zones of semi-invariant resolving power have a complicated shape. We can conclude that there are several zones where the system can resolve objects with a nice finesse. However, various spatial frequencies can be discriminated or, on the contrary, they are resolved impartially due to the action of the axicon illumination.

As the next step, we fixed the distance between the illuminating and imaging elements and compared the experimental results with numerical estimates; see Fig.~\ref{fig:Bessel}~(c).

The contrast, resolution and depth of field of this configuration was evaluated by scanning the aforementioned target (Fig.~\ref{fig:Bessel}). In the first case, the \ac{THz} radiation passes through the diffractive Bessel element (B), which forms a Bessel beam. The latter illuminates the target and \ac{ZP} performs the imaging. During the experiment, the target was scanned several times at different $z$ positions, varying by a $2$~mm step. The best recorded spatial resolution is $1.4 \lambda$. For other distances, when the target is closer or farther away from the axicon, the spatial resolution varies from $2.4 \lambda$ to $1.6 \lambda$. This experiment enables us to evaluate the \ac{DOF} of the setup around a particular "focal spot". In general, the experiment coincides with numerical simulations, shown in Fig.~\ref{fig:Bessel}~(g). The groups of different numbers of bars per millimeter are different in both the experiment and the theory.

Now we switch an axicon to \ac{ZP} and repeat the measurement with a finer mesh of $z$ values (the step size is now 1~mm) (Fig.~\ref{fig:Bessel})~(d). We observe immediate changes in the MTF - the high-contrast areas are now continuous, and they are not separated anymore by the areas with low contrast. This is a direct indication of the fact that the axicon collects light from its Bessel zone. However, the best resolution is achieved when the axicon is located near the object ($z_2=3$~mm) in single-pixel imaging (marked with a blue arrow); see (Fig.~\ref{fig:Bessel})~(e). As the axicon is displaced further away from the object, the modulation transfer map changes - the low spatial frequencies are still resolved reasonably well, but intermediate spatial frequencies are resolved worse. In particular, high spatial frequencies up to 1.7~lin/mm (denoted by the blue dashed line)  are resolved more or less homogeneously without pronounced changes. We note that the optimally resolved spatial details are up to the spatial frequency of the light-collecting axicon ($\xi =0.78~\text{mm}^{-1}$).

The modulation transfer image reveals quite an intriguing feature of the axicon light-collecting system -- it does have a very extended \ac{DOF}, see (Fig.~\ref{fig:Bessel})~(e). This imaging element has an outstanding \ac{DOF} length, which was already implemented to observe samples with varying axial structure \cite{minkevicius2019bessel}.

This behavior was verified both numerically and experimentally using the sample introduced in the previous part of the manuscript; see Fig.~\ref{fig:Bessel}~(f). A comparison of the experimentally obtained \ac{MTF} with the numerically calculated \ac{MTF} is shown in Fig.~\ref{fig:Bessel}~(g). Experimental estimates of contrast, resolution and depth of field are in line with numerical expectations.

The irradiance is represented (both in experimental and numerical terms) in Fig.\ref{fig:Bessel}~(h). The highest brightness of the images recorded was observed at distances different from those predicted by the modulation transfer map, compare Figs.~\ref{fig:Bessel}~(b,e) and (h).
In this case, the results show better quality when the Bessel beam is formed directly into the detector. The spatial resolution of the best image is $1.2 \lambda$. Meanwhile, images scanned around the maximum intensity point have a much better spatial resolution that varies from $1.6 \lambda$ to $1.4 \lambda$.

\subsection{\label{sec:single_pixel_imaging}Single-pixel imaging using the cubic phase mask}

\begin{figure}
    \centering
    \includegraphics[width=1\columnwidth]{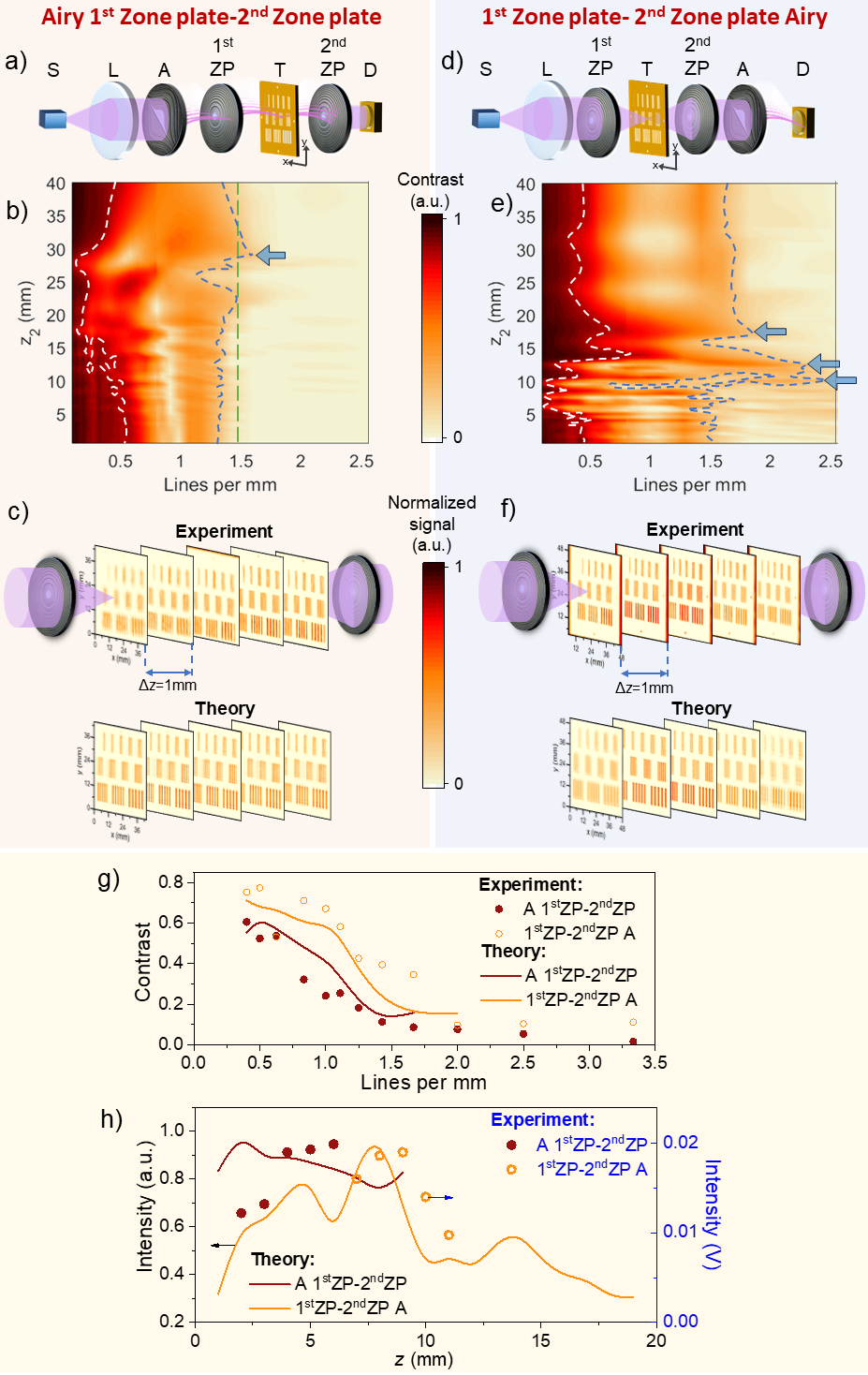}
    \caption{Target imaging using the cubic phase mask and a diffractive zone plate in two different setups: panel (a) -- when the Airy mask serves as an illuminating and the zone plate as a collecting element, (d) their reverse combination. In both setups: S is a \ac{THz} source, L denotes the converging lens, A is the Airy phase mask, ZP stands for the zone plate, T is the imaged target, and D is the detector. Panels (b) and (e) present numerically estimated modulation transfer map dependencies on the distance between the target and the second optical element. The white dashed line in both panels (b) and (e) frames the boundary where the contrast value is greater than $80 \%$ of the maximal value. The dashed green line in panel (b) denotes the resolution of $1.5$~lin/mm; dash blue outline marks the boundary of $20 \%$ of the contrast value, the blue arrows highlight areas of a resolution reaching $1.8$~lin/mm, $2.3$~lin/mm and $2.4$~lin/mm in panel (e), here also note the region extending nearly homogeneously from $z_2=40$~mm up to $z_2=40$~mm; panels (c) and (f) depict experimentally obtained (top picture) and numerically estimated (bottom picture) \ac{THz} images of the target. The distributions are normalized to the maximum value of the signal; panel (g) --  the modulation transfer function; panel (h) presents intensity dependence on the distance between the first optical element and the imaged target.}
    \label{fig:Airy}
\end{figure}

A more complicated phase mask can also be introduced into the imaging scenario to create a structured \ac{THz} illumination and take advantage of enhanced resolution, better materials inspection, etc. \cite{ivaskeviciute2022}. These fascinating novel findings motivated us to probe the performance of the cubic phase mask in the sample illumination and imaging scenarios, see Fig.~\ref{fig:Airy}.

We first consider the scenario, when the incoming \ac{THz} wave hits the Airy phase mask, which is placed at a distance of $1$~cm from the first \ac{ZP}; see Fig.~\ref{fig:Airy}~(a). Thus, the structured \ac{THz} Airy illumination is created for the target inspection. The light scattered by the target is collected by the second \ac{ZP} and directed to the single-pixel detector. We place the object at the spot where the Airy beam is the brightest and numerically estimate the contrast map for different positions of the light-collecting zone plate from the target under inspection. This setup was already discussed in the literature \cite{ivaskeviciute2022}, therefore, we are curious whether any further improvements can be introduced.

Numerical experimentation predicts strikingly better imaging: for most positions of the light-collecting imaging element, up to $1.5$~lin/mm can be resolved. Of course, there are two distance ranges where the resolution is poor, the region around $z_2 \sim 18$~mm and around $z_2 \sim 25$~mm. However, for distance $z_2 \sim 29$~mm, the resolution increases to $1.5$~lines per millimeter marked with a blue arrow in Fig.~\ref{fig:Airy}~(b). Compared to the case discussed in the previous section, this is a noticeable increase. The light collection is carried out using a zone plate; therefore, we estimate that $z_1^{-1}+z_2^{-1} = 0.121~\text{mm}^{-1}$ gives the effective $f = 8.26$~mm, which is less than the actual focal point distance of $8.5$~mm. We also note that the modulation transfer map indicates a better resolution power for objects with low spatial frequencies, and the response to intermediate frequencies demonstrates a complicated "snake"-like dependency. High spatial frequencies are better resolved in one particular region denoted by a blue arrow, see Fig.~\ref{fig:Airy}~(b).

We recall the third metric, which is crucial in the evaluation of imaging systems, the \ac{DOF}. First, we observe three distinct regions with their \ac{DOF}; see Fig.~\ref{fig:Airy}~(b). One is located very close to the element and extends up to $z_2 \sim 15$~mm. The second, around $z_2 \sim 25$~mm, is directed towards the element. The next is located at $z_2 \sim 40$~mm. 
Yet another property is that the modulation transfer function has a distance region when the resolving power decreases and increases afterward, see Fig.~\ref{fig:Airy}~(b, e). This makes the discussion in terms of \ac{DOF} troublesome.

Subsequently, as in previous cases, during the experiment, the target is moved in the $z$ direction with a 1~mm step to experimentally estimate contrast, resolution and depth of field. The recorded images are presented in Fig.~\ref{fig:Airy}~(c) together with the numerically obtained results. Generally, the agreement is on a high level. When comparing results to those of the previous Section, the combination of the Airy lens with \ac{ZP} allows for better images - they are now clearer and less distorted even for the target moved out from the focal point. The spatial resolution at the maximum intensity point is $1.4 \lambda$.

In the next experiment, the target is illuminated by \ac{ZP} but the sample is imaged using a combination of the Airy mask and a \ac{ZP}; see Fig.~\ref{fig:Airy}~(d). Unexpectedly, this change of the elements has resulted in better images of the target for some positions of the light-collecting element! The contrast map for different positions of the elements that produce the image of the sample is shown in Fig.~\ref{fig:Airy}~(e). The main distinction in comparison to the previously discussed combination of illuminating and imaging elements is that regions with good contrast are no longer so freely available. We could determine only a few such regions. The closest one is around $z_2=11$~mm, where the resolution goes up to $2.4$~lin/mm! Slightly further away, at a distance of around $z_2=14$~mm, a resolution of up to $2.3$~lin/mm can be achieved. 
There is a third region around $z_2=18$~mm, where the modulation transfer for intermediate frequencies is good, reaching about $1.8$~lin/mm. As can be seen, the region extends nearly homogeneously up to $z_2=40$~mm. It should be noted that the illumination here is provided through the standard \ac{ZP}, so the rich and distinct map of modulation transfer is caused by the presence of the additional cubic phase plate in the light collection part of the setup. A good resolution for low spatial frequencies is available only for the selection of the sample positions. At some distances, {between $z_2=5$--$10$~mm, even large objects cannot be resolved, see Fig.~\ref{fig:Airy}~(e).

The modulation transfer dependence allows us to estimate the \ac{DOF} in this particular case, see Fig.~\ref{fig:Airy}~(e). Here, we predict a particularly long \ac{DOF}, starting from the axial position of $z_2=14$~mm and extending up to $z_2=40$~mm and beyond. As mentioned previously, some nonuniformity can be observed in the regions with best-resolving power - larger objects are resolved worse, but the finer smaller things can be detected particularly well.

We can assume that \ac{ZP} is the main element in the light collection scheme that provides imaging capabilities; therefore, we estimate $z_1^{-1}+z_2^{-1} = 0.159~\text{mm}^{-1}$ and the effective focal length is $f = 6.28$~mm. As this number quite differs from the expectation, it seems that the imaging capabilities are provided by a proper combination of two elements.

An actual experiment was performed to verify this behavior; its comparison with numerical expectations is given in Fig.~\ref{fig:Airy}~(f). Compared to the case where the Airy lens is positioned in front of the target, placing this element at the rear of the imaging system, in front of the detector, gives better images of a target. The cause for that is the imaging setup, as such behavior for this range of distances is not observed in the two-zone plate system. Thus, Airy phase masks provide better spatial resolution at the highest intensity spot $1.2 \lambda$. 

A comparison of numerically and experimentally determined \ac{MTF} is given in Fig.~\ref{fig:Airy}~(g). It is seen that the experiment and the numerics are in reasonable agreement and show a similar general trend.

The brightness of the recorded image is also an important question, as we have previously determined that the conditions for the best resolution and image brightness are decoupled from each other in previous cases; therefore, we expect here a similar behavior. We now fix the arm of the imaging system and move the target through it to examine the intensity of the recorded image, see Fig.~\ref{fig:Airy}~(h). For the first setup, the intensity drop is observed when the target is closer to the illuminating lens, and this phenomenon is not observed as well in the numerical estimates. However, in the second scenario, an overlap of the experimental results with the numerical ones is seen. Once again, this finding means that if we determine the best position for the sample, using an expectation from the classical single-shot imaging that the image is the brightest at the same place, where the resolution is the best, we miss the spots in the single-pixel imaging with the best-resolving power.

\begin{figure*}
    \centering
    \includegraphics[width=0.8\textwidth]{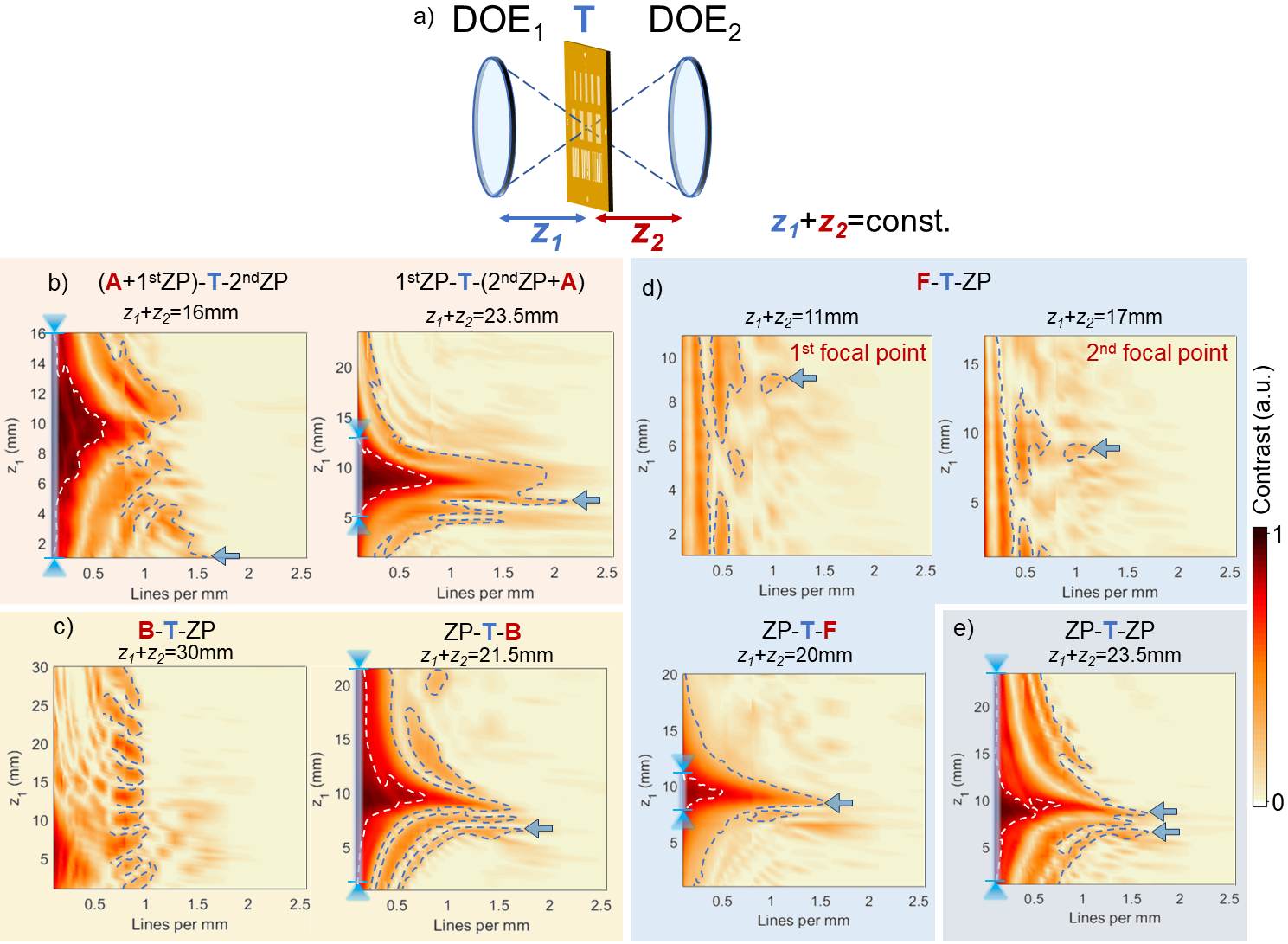}
    \caption{Numerically estimated 2D maps of modulation transfer maps dependence on the target position. (a) setup, where the distance between two diffractive elements (DOE$_{1}$ and DOE$_{1}$) is constant, and only the position of the imaged target changes. 2D modulation transfer maps for two combinations, where (b) Airy, (c) Bessel, (d) Fibonacci, and (e) ZP functions as an illuminating element (left picture) and as a collecting element (right picture). In panel (b), the Airy lens is a combination of the Airy phase mask and the ZP. In panel (d), there are two modulation transfer maps in which the Fibonacci lens as an illuminating element produces two focal spots (top panels), and collects the THz light (bottom panel). Note the absence of a uniform area with good resolution in the upper panels and a pronounced resonance-shaped contrast area reaching a resolution of $1.5$~ lin / mm. Panel (e) -- the performance of the conventional imaging setup with two ZPs for comparison -- note the extended \ac{DOF} reaching 22~mm and sharply expressed resolution reaching nearly $1.7$~lin/mm at 10~mm of $z_2$ distance. Blue stripes and arrows along the $z$ axis indicate extended \ac{DOF} areas; dash blue outline marks the boundary of $20 \%$ of the contrast value, and blue arrows denote areas with exceptional resolution.}
    \label{fig:2D kontrastas}
\end{figure*}

\subsection{\label{sec:optimal_sample_position} Optimal sample position in \ac{THz} imaging experiments}

Numerical experimentation provides a solid foundation to support the choices made during the actual \ac{THz} experiment. Initially, our expectation was that the sample would be placed in the spot where the illuminating structured light was the brightest. These spots were detected both numerically and experimentally; see Fig.~\ref{fig:Verifikacija}.

Findings from the previous subsections reveal that the positions of the brightest images in the single-pixel imaging differ from the positions of the best-resolving power, as illustrated in previous sections. Therefore, a natural question occurs whether the best optimal resolution is achieved when the sample is being illuminated by the focused structured light at those particular spots. To answer this question, we made additional numerical estimates. We fixed the distance between the diffractive illuminating element and the diffractive imaging element at values that were used in the experiments; that is, the sum of the distances $z_1+z_2$ is constant now; see Fig.~\ref{fig:2D kontrastas}~(a). During these numerical experiments, we moved the target between the illuminating element and the light-collecting element. In doing so, we estimate the modulation transfer map, which we present as the dependencies on $z_1$ marked with \ac{DOF} as a blue strip and arrows in Fig.~\ref{fig:2D kontrastas}~(b-e).

\begin{figure}
    \centering
    \includegraphics[width=0.8\columnwidth]{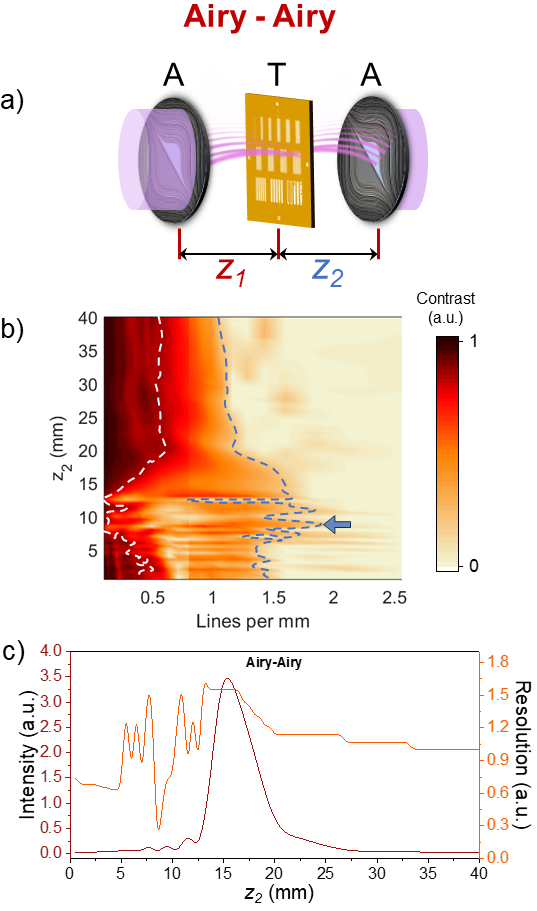} 
    \caption{Numerical estimation of the target imaging using two Airy lenses, both as illuminating and as collecting elements. Panel (a) -- configuration, where the position of the target ($z_1$) is fixed and only the distance ($z_2$) between the target and the second element changes. Panel (b) presents the modulation transfer map on the distance $z_2$, The white dashed outline represents the boundary where the contrast value is greater than $80 \%$ of the maximal value, the dashed blue outline marks the boundary of $20 \%$ of the contrast value, the blue arrow denotes a position with good resolution and contrast. Note the areas nestled below $10$~mm in the $z_2$ scale with a resolution that reaches $1.8$~lin/mm. Panel (c) shows the image irradiance and resolution at distance $z_2$.} 
    \label{fig:A-A}
\end{figure}

For the structured Airy THz illumination, we observe a region where the low spatial frequencies are best resolved, see Fig.~\ref{fig:2D kontrastas}~(b). The intermediate spatial frequencies are well resolved at the position $z_2 \approx 10$~mm where the illuminating Airy beam is the brightest, compared to Fig.~\ref{fig:Verifikacija}~(d). Unfortunately, the high spatial frequencies are worse resolved at this particular position. To resolve higher spatial frequencies, we must increase or decrease the distance from the illuminating element to the target, see the marks in Fig.~\ref{fig:2D kontrastas}~(a). Spatial frequencies up to $1.5$~lin/mm become resolvable in this case, marked with a blue arrow. \textbf{This behavior can be interpreted as the separation of the paraxial action of the system from the nonparaxial action. Lower spatial frequencies correspond to paraxial beams traveling through the effective optical system, whereas higher spatial frequencies are caused by nonparaxial rays propagating from the sample.} Therefore, the conditions when modulation of the fine structures is transferred in an expected fashion could differ from the conditions of paraxial imaging.

The 2D modulation transfer map also reveals here another important metric - the \ac{DOF}, see marks in Fig.~\ref{fig:2D kontrastas}~(b). The zone is placed at $z_2 \approx 10$~mm and extends from the beginning of the parameter space to the very end; thus, we conclude that its length under these conditions is 16~mm.

For structured THz light collection using a cubic phase mask while illuminating with a \ac{ZP}, the modulation transfer map experiences changes compared to the previous case. The position where lower spatial frequencies are resolved with the best power is marked on the graph, see Fig.~\ref{fig:2D kontrastas}~(b). The intermediate spatial frequencies are optimally resolved at the factual focal point of the \ac{ZP} at $z_1 \approx 8.5$~mm. The resolution of those spatial frequencies up to 1~lin/mm is reasonably good when the sample is placed out of focus but closer to the \ac{ZP}, see the part of the map with $z_1 = 6$~mm. This is caused by the asymmetry in the axial profile of the illuminating radiation, see Fig.~\ref{fig:Verifikacija}~(a). As in the previous case, the best distance for high resolution of higher spatial frequencies (around $2$~ lin/mm) is located at a different spot on the modulation transfer map; see the right panel in Fig.~\ref{fig:2D kontrastas}~(b). The most intriguing is the finding that the combination of Airy phase mask and zone plate is expected to provide good resolving power for spatial frequencies up to $2.2$~lin/mm at some particular position indicated by the blue arrow. Of course, since this is the spot where the brightness of both the illuminating beam and the recorded image decreases, a reasonable question arises as to how sensitive the detection system should be. 

The switch of illuminating and imaging elements also results in the decrease of \ac{DOF}, the length of a triangularly shaped zone is now approximately 10~millimeters.

The next case involves the Bessel axicon and the zone plate; see Fig.~\ref{fig:2D kontrastas}~(c). The axial intensity distribution of the Bessel beam is a focal line with pronounced intensity oscillations. This behavior is somewhat translated to the modulation transfer map, as for low spatial frequencies, we observe distances where those frequencies are optimally resolved and where they are not. However, it is worth underlining that these positions are not always positions where the intermediate and high spatial frequencies are clearly resolved, i.~e., the maximal values of the modulation transfer map are not horizontal lines, but rather a complicated set of shapes. The Bessel beam corresponds to a conical nondiffracting beam created by plane waves with a wavevector lying on the cone with one particular frequency. In our case, this was the angle of $\beta =0.4$ rad. This fact directly translates to the increase in resolution for spatial frequencies around 0.75~lin/mm; see Fig.~\ref{fig:2D kontrastas}~(c). It is apparent that higher spatial frequencies are resolved only in a few particular spots; see, for instance, the locations at $z_1 \in (2, 13)$ millimeters, and spatial frequencies up to $1$~lin/mm are resolved with a sufficiently large contrast.

The definition of \ac{DOF} is somehow troublesome for the case of axicon illumination because of the non-uniform nature of the modulation transfer function. However, if we stick to the definition that requires high contrast values, it can be defined as stretching from the element to the $z_1=12$~mm, thus being comparable in length to the two previous cases.

When the axicon is used for light collection and the sample is illuminated by a zone plate, the low spatial frequencies corresponding to the paraxial rays are optimally resolved when the intensity of the illuminating beam is the highest at $z_1=8.5$~mm. This situation is especially interesting when we compare it to the previous case where a combination of the zone plate and the cubic phase mask focuses the light on the single-pixel detector. We see that low frequencies are comparatively well resolved for larger ranges of positions. Intermediate frequencies are reasonably well pronounced there when it should not be expected from the standpoint of the paraxial imaging theory; see the area for spatial frequencies up to $1$~lin/mm and note the high contrast is moving further away from the zone plate. Higher spatial frequencies are also resolved at different positions reaching $1.7$~ lin/mm within $7-9$~ mm on the $z_1$ axis. In particular, we see that the axicon provides better resolution than the \ac{ZP}, but it is worse than the Airy phase mask together with \ac{ZP}.

The \ac{DOF} is relatively easy to define for this case, as it is a more or less uniform triangular shape that extends from the element up to 25 millimeters.

We move on to the performance of the Fibonacci element in combination with a \ac{ZP}, see Fig.~\ref{fig:2D kontrastas}~(d). As the element produces two focal spots, we investigate the illumination of the target when placed in both of them. The presence of bifocal spots is easily noticeable for low spatial frequencies up to 0.5~lin/mm. As the spatial frequencies in the sample increase, they are no longer resolved at the same positions as the lower spatial frequencies, but they are resolved at different distances from the light-illuminating element. The dynamics are more complex as the rays in the system become more and more nonparaxial, we see the appearance of numerous positions where they can still be resolved, and resolution can reach $1$~lin/mm at the particular place marked by the blue arrow. Most importantly, placing the object in the first focal point seems to be less optimal than placing it in the second focal point, since the target should become blurred. There is no distinct zone with a classical depth of field for the Fibonacci illuminating element, though one can argue that it is still present as a very narrow long zone stretching along the $z_1$ axis.

\begin{figure}
    \centering
    \includegraphics[width=0.8\columnwidth]{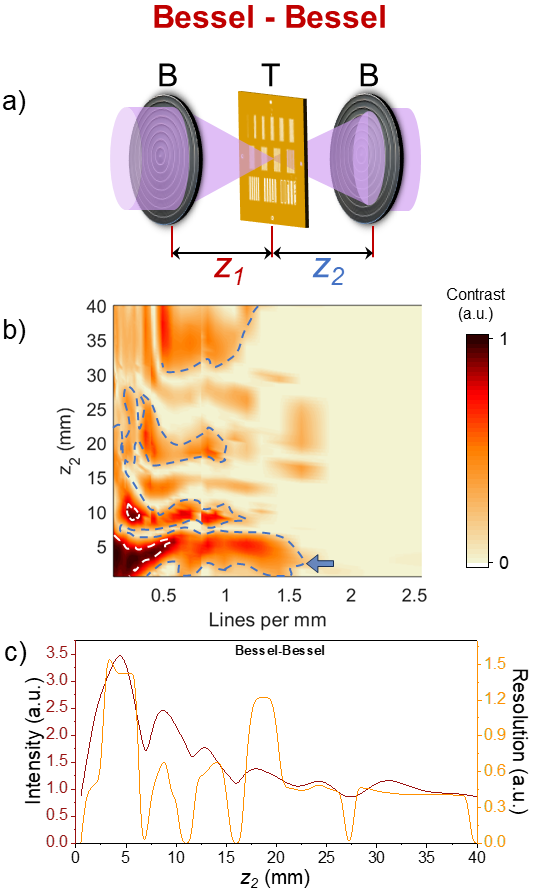} 
    \caption{Numerical estimation of the target imaging using two Bessel axicons, both as illuminating and as collecting elements. Panel (a) shows the configuration, where the position of the target ($z_1$) is fixed and only the distance ($z_2$) between the target and the second element is changed. Panel (b) -- dependence of the modulation contrast map on the distance $z_2$. The white dashed outline represents the boundary where the contrast value is greater than $80 \%$ of the maximal value, the dashed blue outline marks the boundary of $20 \%$ of the contrast value, the blue arrow denotes positions with $1.6$~lin/mm resolution and good contrast. Panel (c) presents intensity and resolution dependence on the distance $z_2$.} 
    \label{fig:B-B}
\end{figure}

When the zone plate is used to illuminate the sample, the optimal distance for the low spatial frequencies seems to coincide with the spot with the brightest irradiation. However, the region of distances when paraxial objects should be resolved has a maximum shift further away from the focal spot, see the 3$^\text{rd}$ inlet in Fig.~\ref{fig:2D kontrastas}~(d).

As we move on the modulation transfer map to the intermediate frequencies, we observe an appearance of new regions, where these objects can be resolvable. This behavior culminates in the appearance of distances when nonparaxial objects become resolvable. In this case, we still have a clearly pronounced zone of the classical \ac{DOF}, which is approximately 5 millimeters long.

Lastly, for comparison, we check the behavior of the conventional imaging setup with two \acp{ZP}; see Fig.~\ref{fig:2D kontrastas}~(e). In a quite expected fashion, the position of the brightest illumination results in the position of the best-resolving power for low spatial frequencies. We note that on the z-axis there are a few other areas where paraxial objects should be resolvable with slightly worse contrast, which is a distinction of the single-pixel imaging from the single-shot imaging; see Fig.~\ref{fig:2D kontrastas}~(e). Intermediate spatial frequencies are well resolved at approximately the same position as the focal point -- it can be seen by expressed resolution reaching nearly $1.5$~lin/mm around 10~mm of $z_2$ distance. It can be seen that higher spatial frequencies are better resolved in some particular areas closer to the illuminating \ac{ZP}. In the last case, \ac{DOF} can be estimated to extend up to the range of $22$~mm. 

Thus, our numerical experimentation has largely validated our experimental choice of placing the object in the brightest spot after the diffractive element. However, these findings indicate that the condition for the best illumination and the best resolution becomes less dependent on nonparaxial high spatial frequency components of the image. Given difficulties in the experimental implementation, when deviations in the actions of the actual element from the intended might appear, this triggers the question of how difficult it is to obtain high resolution under realistic experimental conditions.

\begin{figure}
    \centering
    \includegraphics[width=1\columnwidth]{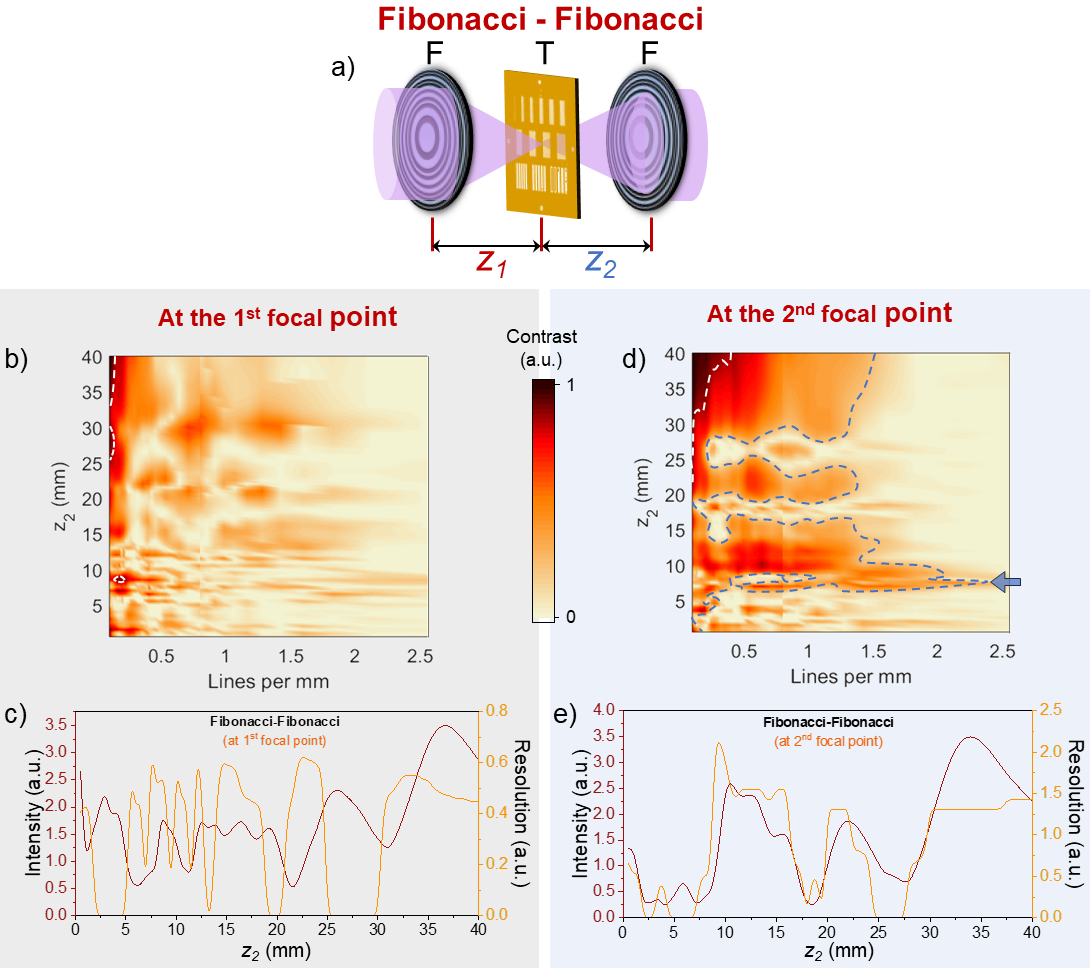} 
    \caption{Numerical estimation of the target imaging using two Fibonacci lenses, both as illuminating and as collecting elements. Panel (a) depicts the setup, where the position of the target ($z_1$) is fixed and only the distance ($z_2$) between the target and the second element changes. As the Fibonacci lens produces two focal points, the results of the first focal point are depicted in panel (b), and the results of the second focal point are shown in panel (d). White dashed outline represents the boundary where the contrast value is greater than $80 \%$ of the maximal value, the dashed blue outline marks the boundary of $20 \%$ of the contrast value, the blue arrow denotes an area with exceptional resolution reaching $2.4$~lin/mm. Panels (c,e) show intensity and resolution dependence on the distance $z_2$.} 
    \label{fig:F-F}
\end{figure}

\subsection{\label{sec:single_pixel_imaging_using_doublets}Nonconventional single-pixel imaging using various doublets of zone plates, Bessel, Airy, and Fibonacci elements}

The last part of the investigation is purely numerical in order to derive principles of the nonparaxial \ac{THz} images relating light illuminating/collecting elements and their optimal placement for the rational setup design. Now, when we understand how the position $z_1$ of the target with respect to the illuminating element influences the resulting modulation transfer, for the sake of brevity, we select these distances in our further numerical experimentation and investigate how the position $z_2$ of the diffractive imaging element influences the details resolved in the target.

We start with a setup involving a doublet of \acp{ZP} in combination with Airy phase masks, see Fig.~\ref{fig:A-A}~(a). The modulation transfer map for various positions $z_2$ of the light-collecting element is shown in Fig.~\ref{fig:A-A}~(b). 

First, we notice an extended range of positions for the imaging element to be placed with respect to the object since the \ac{DOF} of this setup is large. The most optimal range distances for resolving low-frequency details are for positions of $z_2$ starting from 20~mm up to 32~mm. The situation is more or less the same for intermediately low spatial frequencies, with the best contrast locations closer to $z_2= 20$~mm. 

However, as the spatial frequencies in the object become more and more nonparaxial, conditions for the best resolution change, with the distance $z_2= 20$~mm ensuring a good resolution of around $1.3$~lin/mm for both large and small details. However, a better resolution is expected for details in the nonparaxial range, for example, at $z_2= 10$~mm and reaching $1.8$~lin / mm. 

Throughout our review of these imaging benchmarks, we did find an indication that in single-pixel imaging, the image quality assessment is not as straightforward as it is in single-shot imaging. Conditions for the best image brightness and best resolution are mutually opposing and place objects, illuminating and light-collecting diffractive elements at different positions. The same holds, and for the setup under investigation, see Fig.~\ref{fig:A-A}~(c). The particular position in which the image is brightest is different from that in which the resolution is better. Moreover, in the areas where the image is not bright anymore, a reasonably good resolution is possible.

In the next experimentation, we turn to a doublet of axicons, see Fig.~\ref{fig:B-B}~(a). The modulation transfer map is depicted in Fig.~\ref{fig:B-B}~(b). Here, we see a distinct feature of this setup, the modulation transfer map contains a few areas where the low, intermediate, and high spatial frequencies are resolved at the same time; see marked areas in the plot. The presence of the fine structure within those areas of the modulation transfer map indicates that an optimal choice is to place the second axicon closer to the target, which will give a resolving power to the setup of up to $1.6$~lin/mm.

As expected from the investigation of the aforementioned setups, the conditions for image brightness and image clearness are different, although a slight correlation can be observed between the spots where the image is bright and the spots where the sample is resolved; see Fig.~\ref{fig:B-B}~(c).

Finally, we explore the imaging using two Fibonacci lenses; see Fig.~\ref{fig:F-F}~(a). For this setup, we have two possible positions of the target, either to be placed in the first focal spot or in the second, for which we calculate modulation transfer maps; see Fig.~\ref{fig:F-F}~(b,d). A comparison of these maps reveals that the choice of the first focal point results in suboptimal resolved spatial frequencies, as there is no homogeneous region where all frequencies can be resolved at the same time, see Fig.~\ref{fig:F-F}~(b). Furthermore, a comparison of the image brightness and resolution, see Fig.~\ref{fig:F-F}~(c), indicates difficulties in the experimental implementation of this setup, as the brightest spots usually do not coincide with the places where the structure of the sample can be resolved.

The situation improves, as the target is illuminated by the second focal point of the bi-focal lens, see Fig.~\ref{fig:F-F}~(d). Though there are sample positions where the Fibonacci lens resolves neither paraxial spatial frequencies nor nonparaxial ones, there are a few positions where they can be simultaneously resolved. This is especially promising for one particular position $z_2 = 9$--$10$~mm; see Fig.~\ref{fig:F-F}~(d), where up to $2.4$~lin/mm is expected to be resolved!

As the comparison in the brightness of the image with the resolution of the optical system reveals, see Fig.~\ref{fig:F-F}~(e), this particular distance is also close to the spot, where the image should be the brightest. This finding means that an experimental implementation of such an imaging system should not be ruled out from a practical point of view.

\begin{figure}
    \centering
    \includegraphics[width=0.96\columnwidth]{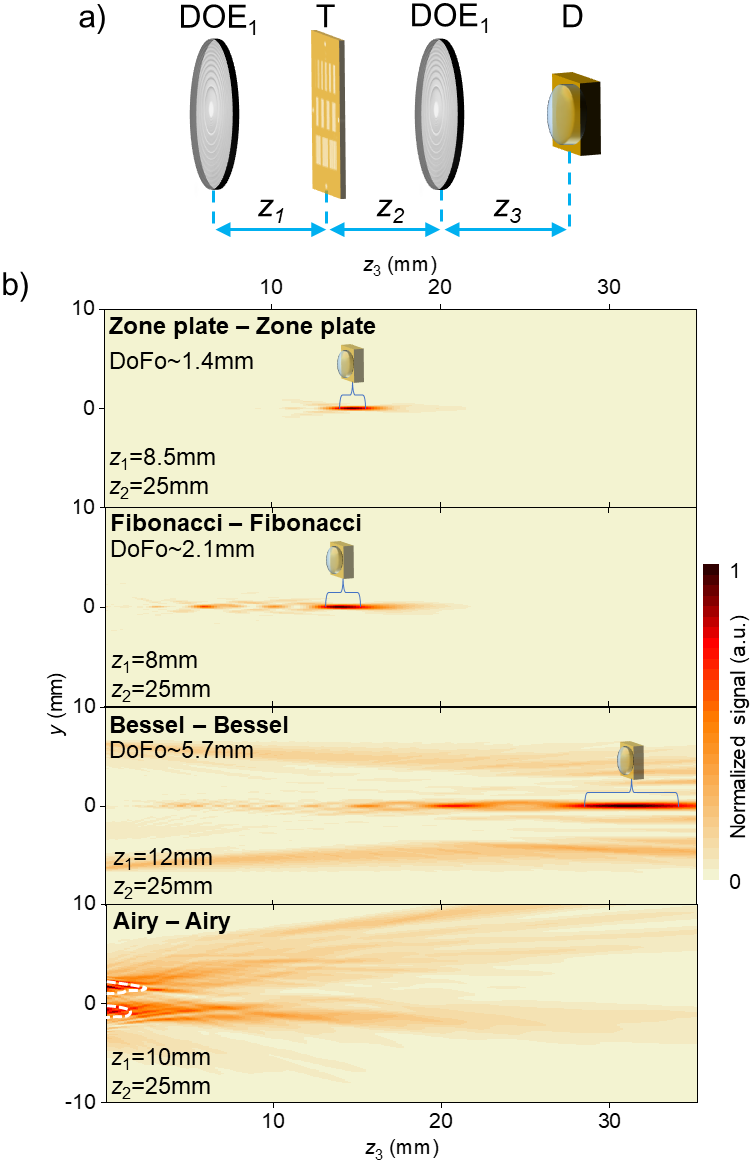}
    \caption{Numerically estimated \acf{DoFo} in schemes with symmetric illumination and collection of THz light. Panel (a) -- optical setup for the evaluation of \ac{DoFo}. DOE$_1$ marks two identical diffractive optical elements, T is the imaged target and D is the detector. Panel (b) presents the intensity distribution versus the distance $z_3$ between the collecting optical element and the sensor plane at the most optimal distances $z_1$ between the illuminating element and the target and $z_2$ between the imaged target and the collecting element. Four different symmetric illuminating/collecting schemes: Fresnel zone plates (top panel), Fibonacci lenses (second from top), Bessel axicon (third from top), and Airy zone plate (bottom panel). The insets represent the DoFo -- the region where the detector can capture the sharpest image. The DoFo corresponds to the area where the intensity exceeds $80 \%$ of the maximum signal. In the case of the Airy -- Airy zone plates combination, the area where the intensity exceeds $80 \%$ is represented by a white dashed outline. The colored scale represents the normalized signal intensity.} 
    \label{fig:DOF}
\end{figure}

\section{On the depth of focus in THz imaging setups}

When the concept of resolution is discussed in optical microscopy, the primary focus is typically on lateral resolution, which concerns the ability to distinguish points in the plane perpendicular to the optical axis. Another critical element of resolution is the axial (or longitudinal) resolving capacity of the objective, which relates to the measurement of depth along the optical axis and is commonly referred to as "Depth of Field", \ac{DOF}. Just as in conventional photography, the \ac{DOF} is determined by the range from the nearest in-focus object plane to the furthest one simultaneously in focus. We have already discussed this in the previous sections.
A similar concept is the \acf{DoFo}}. The \ac{DoFo} defines the range of distances from the sensor plane to which the image projected by the lens is acceptably sharp. Note that the term "depth of focus," which relates to the image space, is often used interchangeably with "depth of field," which relates to the object space. This interchangeable use of terminology can lead to confusion, especially when both terms are used to specifically refer to the depth of field in microscope objectives. Depth of focus is also not a fixed value but rather a range of acceptable sharpness that varies depending on the viewing conditions and the desired image quality. The main difference between the "depth of field" and the "depth of focus" is that depth of field refers to the object space, or the quality of the image coming from a stationary lens as the object is repositioned, whereas depth of focus refers to the image space, or the ability of the sensor to maintain focus for different sensor positions, including tilt.

\begin{figure*}
    \centering
    \includegraphics[width=0.9\textwidth]{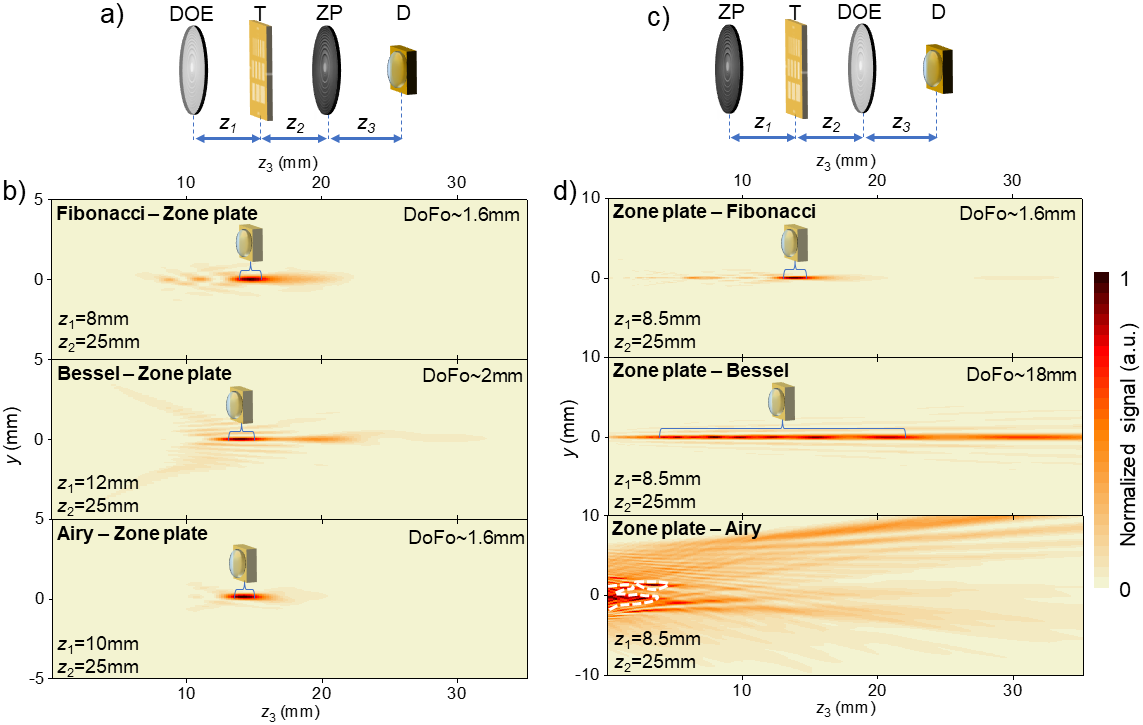}
    \caption{Numerically estimated depth of focus (DoFo) in schemes with asymmetric illumination and collection of THz light. Panels (a) and (c) -- optical setups for evaluation of the DoFo containing diffractive optical elements for THz light illumination and ZP for light collection and inverse arrangements in the setup, respectively. DOE$_1$ marks the diffractive optical element investigated, ZP denotes the Fresnel zone plate, T is the imaged target, and D is the detector.  Panels (b) and (d) present the intensity distribution versus the distance $z_3$ between the collecting optical element and the sensor plane at the most optimal distances $z_1$ between the illuminating element and the target and $z_2$ between the imaged target and the collecting element.  Three different asymmetric THz light illuminating/collecting schemes in (b): Fibonacci lens -- ZP (top panel), Bessel axicon -- ZP (middle panel), and Airy zone plate -- ZP (bottom panel).  Panel (d) presents intensity distributions using the inverse arrangement of the set-up, that is, ZP serves as an illuminating element, while other DOEs serve as light-collecting components. The insets demonstrate the DoFo, which represents the region in which the detector can capture the sharpest image. The DoFo corresponds to the area where the intensity exceeds $80 \%$ of the maximum signal. In the case of the conventional ZP -- Airy zone plate combination, the area where the intensity exceeds $80 \%$ is represented by a white dashed outline. The colored scale represents normalized signal intensity.} 
    \label{fig:DoFo_asimetrine}
\end{figure*}

As an illustration of this difference, it can serve the following considerations: Imagine taking a picture of a flower with a camera. If one moves the flower closer to or farther from the lens, it will change the depth of field, or how much of the flower and its surroundings are in the focus. If you move the sensor inside your camera closer to or farther from the lens, the depth of focus will change or, in other words, how much of the image projected by the lens is in focus.

In what follows, we extend similar examinations to different compact THz imaging scenarios and estimate the depth of focus in our numerical virtual experiments using the setup depicted in Fig.~\ref{fig:DOF}~(a) and  Fig.~\ref{fig:DoFo_asimetrine}~(a,c). Two imaging approaches were considered: the first is when the THz light illuminating and collecting optical schemes are symmetric, that is, they use the same optical elements for THz light illumination and its collection, see Fig.~\ref{fig:DOF}~(a). The second case is asymmetric, that is, when the different diffractive elements for the engineering of THz light and its collection are used in combination with the conventional silicon ZP, see Fig.~\ref{fig:DoFo_asimetrine}~(a,c). As the main aim of these estimates is to optimally position the single-pixel detector, we place the target at the distance $z_1$ from the illuminating element and at the distance $z_2$ from the collecting element in all cases. We are interested in the electric field distribution in the transverse plane after the collecting element. 

The first scenario is the conventional setup consisting of two ZPs, see top panel in Fig.~\ref{fig:DOF}~(b). In this case, the DoFo is a point-like region, located at a distance of around $z_3=14$~mm from the collecting element and is approximately $1.4$~mm long. This numerical estimate is approximately in line with the experimental value obtained in the laboratory -- the distance was $\approx 11.4$~mm. Given that the single-pixel sensor had an additional hemisphere attached to it, which should extend the actual propagation length by a few millimeters, the overall agreement is as expected.

In the second scenario, Fibonacci lenses perform both light collection and illumination tasks, see the second from the top panel in Fig.~\ref{fig:DOF}~(b). 
From the intensity distribution, immediately we note the distinct feature of the bifocal element, the presence of two focal spots together with secondary axial oscillations. The brightest focal point is the second, which collects light at a distance of $z_3 \approx 13$~mm, and the DoFo extends approximately $2.1$~mm around that spot.

We move on to the Bessel illumination and the axicon collection of THz light; see the third panel in Fig.~\ref{fig:DOF}~(b). In this case, we observe the characteristic properties of the conical phase element, where it focuses the light coming from the target in a conical fashion. Keep in mind that phase masks that illuminate and perform focusing are those of the axicon with cone angle $\beta = 0.4$~rad. The characteristic size of the element $D$ is approximately 1~inch (that is, $D=25.4$ mm), so we expect the conical radiation to converge to the brightest focal spot at a distance approximately equal to $z_3=D/(2 \sin \beta) \approx 32$~mm. This results in a notable feature of the axicon -- extended long focal line, which is also present in the simulation in both the illumination and the collection. The sum of distances $z_1+z_2 = 37$~mm, so without the target in the setup, the second axicon would collect the light that exits the Bessel zone of the focal line. The presence of the target changes the angular frequencies of the THz light reaching the collecting axicon, see Fig.~\ref{fig:DOF}~(b). The intensity of the electric field coming from the target obviously has a ring-like shape with some field present also in the center. The conical nature of the second element is clearly revealed here. First, due to the geometry of the DOE, which comprises a circular diffractive grating with a period proportional to $k \sin \beta$, the axial electric field linearly increases its amplitude. This effect can be understood by thinking that each circular grating at distance $r$ from the center of the element has a surface area that is proportional to $2 \pi r$, so purely geometrically this translates to an increase in the electric field when we move on the focal axis away from the axicon. We note here that alternative ways of generating a Bessel beam can result in a constant axial intensity profile. The second effect, which is observed here, is caused by the fact that the actual element is a multi-level DOE. As the phase profile is not smooth, we observe an appearance of the axial oscillations, see the third panel in Fig.~\ref{fig:DOF}~(b). We estimate the DoFo to be approximately $5.7$~mm long.

The last symmetric case, we consider here, is the Airy beam performing illumination tasks and the cubic phase mask (the Airy lens) performing THz light collection; see the bottom panel in Fig.~\ref{fig:DOF}~(b). This case is the most intriguing, as the Airy phase mask should move the focal spot away from the axis. In the simulation, we observe two distinct inhomogeneous zones where the collected THz light is the brightest. One is located above the optical axis, and the second is slightly below it and extends towards it as the light propagates. Thus, keeping the discussion consecutive, we indicate two regions where the DoFo can be defined. However, the detector in our numerical simulations was located at a position of $z_3=11.4$~mm, which corresponds to the case where the Airy element focused the radiation on the axis at this particular position. This indicates that imaging can still be performed even if the detector is placed in a suboptimal position. As our previous discussion indicated, the resolution, contrast, and depth of field were not negatively affected by the placement of the detector. The most notable effect of this choice was the decrease in the brightness of the recorded image. This finding raises the question of whether the location of the optimal focal spot and the DoFo is essential in single-pixel imaging using structured light.

The results of the investigation using asymmetric setups are presented in Fig.~\ref{fig:DoFo_asimetrine}. As can be seen, the illuminating element has a distinct effect on the distribution of the intensity of the electric field after the collecting element. Let us start with the setup, where the ZP performs the direction of light to the detector, while the illumination is structured; see Fig.~\ref{fig:DoFo_asimetrine}~(b). The most common feature is that the position of the brightest spot is approximately always constant and seems to be unaffected by the illumination -- it is approximately at $z_3=14$~mm, which more or less agrees with actual experiments and once again supports the choice of the detector position. However, in all cases, we observe some distinct fine features.

For illumination using a Fibonacci lens, we still observe a few distinct spots before the brightest one, hence, in principle, one could place a single pixel at locations closer to the collecting element, see Fig.~\ref{fig:DoFo_asimetrine}~(b). We estimate the DoFo to be approximately $1.6$~mm.

The illumination of the Bessel beam is of a conical nature, so even after the target is placed in the Bessel zone, the intensity profile after the target still has a ring-shaped shape of the signal when it reaches the ZP that directs the light to the focal spot; see Fig.~\ref{fig:DoFo_asimetrine}~(b). As the THz light incident on the ZP has a ring-like shape, we observe a formation of the Bessel-like focal spot after the ZP. Note the conical arms reaching the focal spot and the presence of additional fringes in the focal plane. The extended focal line is also present there. The DoFo is also larger here, it is up to $2$~mm in length, although the actual region of the high axial intensity is approximately $10$~mm. This suggests that moving the detector might be worth considering.

The Airy beam illumination results in a rather compact intensity distribution in the longitudinal plane, see Fig.~\ref{fig:DoFo_asimetrine}~(b). The area of the highest intensity appears to be slightly shifted from the optical axis, but the optimal depth of focus region covers the optical axis. We estimate the DoFo to be approximately $1.6$ mm.

Switching the illuminating element with the light collecting results in different positions for the brightest spot, see Fig.~\ref{fig:DoFo_asimetrine}~(d). First, for the collecting Fibonacci element, we observe the appearance of two distinct focal spots, the second one being the brightest. The location of the brightest spot is approximately $14$ mm and its extent is $1.6$~mm.

The axicon element is supposed to collect the incident light on a long focal line, which is the case observed at the next inlet; see Fig.~\ref{fig:DoFo_asimetrine} (d). The light reaching the axicon element from the target illuminated with a ZP is more or less homogeneous around the center of the element, so the focal zone of the axicon DOE is clearly shaped. This indicates that there is a wide range of locations on the axis available for detector placement. The extent of DoFo is estimated to be $18$~mm.

The Airy phase mask in conjunction with a ZP can also perform light collection tasks; see the bottom panel in Fig.~\ref{fig:DoFo_asimetrine}~(d). The intensity distribution in the longitudinal plane is inhomogeneous, with bright spots located rather close to the DOE. However, as our previous findings have indicated, moving the detector further away from these positions results in promising imaging results.

\section{Discussion and Conclusions}

Most common single-pixel imaging schemes involve basis scan or compressive sampling \cite{duarte2008single}, where structured illumination or structured detection is ensured using Hadamard matrices, Fourier basis, or random patterns \cite{gibson2020single}. In these setups, the sample is placed in a fixed position and does not move in the transverse plane. An alternative to these popular approaches is the so-called raster scanning of the sample in the transverse plane. This approach is easier to implement, as it does not require additional computational resources to perform image retrieval. However, it has some disadvantages when compared to those more computationally intense methods, see Ref.~\cite{duarte2008single}.

From a theoretical point of view, the raster scan method is pretty straightforward, as the test functions are paraxially assumed to be delta functions and thus the measurement directly provides the object. The findings of this work indicate that this is not the case anymore in the nonparaxial regime, since the introduction of structured illumination and structured detection into the raster scanning scheme results in distinct and observable differences in contrast, resolution, depths of field, and focus. Moreover, the optimal placement of diffractive optics illuminating and detecting elements is also affected by the nature of these elements.

The importance of THz structured light illumination and its collection using nonparaxial diffractive optical elements in THz imaging can be illustrated by a comparison of the benchmarking numbers. 
\textbf{In the case of nonparaxial structured THz light illumination,} when the conventional zone plate collects light, the Airy beam illumination enables a nearly homogeneous resolution of around $1.3$~lin/mm in the modulation transfer map, varying the distance between the target and the optical elements within 40~mm range as well as extended depth of field in the same range -- these features are in pointed contrast to the illumination cases of using the Fibonacci lens and the Bessel axicon. 
\textbf{In the case of nonparaxial structured THz light collection,} when the zone plate is used to manage illuminating THz light, and the Bessel axicon collects it, this approach permits the realization of a nearly homogeneous zone in the modulation map with strongly expressed contrast around $0.5$~lin/mm resolution and moderate contrast of $1.7$~lin/mm resolution as well as the depth of field ranging up to 40~mm. One can note that usage of the Airy zone plate and the Fibonacci lens in the light collection scheme empowers pronounced resolution within the range of $2.4-2.5$~lin/mm, however, good contrast areas and depth of field, in particular, in the Fibonacci case, are inhomogeneous in the modulation transfer map. The employment of the Airy zone plate in the THz light collection implements essential smoothing of the depth of field and contrast zone in comparison to the Fibonacci case -- it extends from 17~mm up to 40~mm still keeping here the reasonable resolution of $1.7$~lin/mm.

These findings and results open new insights into the raster scanning technique, as they show evidence that the test functions for each individual setup are unique and are able to revitalize this straightforward technique by bringing advancements to one of the oldest single-pixel methodologies without any additional computational costs of the competing approaches.

In conclusion, a comprehensive study dedicated to the engineering of THz light and the positioning of the components of silicon diffractive optics was carried out in nonparaxial THz imaging. The single-pixel (focused beam) raster scanning imaging strategy that uses structured illumination and detection is introduced and compared to conventional single-shot imaging. Differences between paraxial and nonparaxial compact THz imaging schemes are revealed via numerical and experimental investigations covering a variety of THz light illumination/detection schemes involving symmetric and asymmetric combinations of conventional Fresnel zone plates, Fibonacci lenses, Bessel axicon, and Airy zone plates. Numerical modeling using the Rayleigh-Sommerfeld diffraction integral together with experimental realizations enabled us to find the optimal positions of the optical components in \ac{THz} light illumination/detection schemes, estimate their influence on the properties of images, and unveil the conditions when the optical nonparaxiality in the setup decouples the brightest point in the \ac{THz} image from the best position for qualitative THz image recording.

Throughout the investigation, special emphasis is placed on the effect on the image resolution part of the setup, indicating how the structure of the point-spread function of the system helps improve the imaging of an object. It is methodically benchmarked in various asymmetric and symmetric scenarios in the THz light illumination/collection schemes, and their performance is reported in different metrics such as modulation transfer function, spatial resolution, contrast, depth of field, and depth of focus. The experiments performed are well supported by extensive numerical investigations. 

These extensive investigations formed a broad basis to derive \textbf{principles for diffractive optics-based THz light engineering and optical components location in optical setup enabling rational design in a compact single-pixel nonparaxial THz imaging.} As we observed \textbf{sharp differences a) between single-shot and single-pixel imaging and b) between various single-pixel schemes}, it is of particular importance to introduce principles in defining qualitative image formation conditions for further development of compact THz imaging setups.

\textbf{Therefore, relying on the findings aforesaid, we derive the following principles for diffractive optics-based light engineering and optimal assembly of optical components enabling rational design in a compact single-pixel nonparaxial THz imaging:}

\textbf{First,} in the nonparaxial regime, the single-pixel raster scan THz imaging can be significantly improved in terms of contrast, resolution, and depth of field by the introduction of structured THz light illumination and collection using diffractive optical elements. 

\textbf{Second,} in nonparaxial THz imaging, the optimal modulation transfer condition appears to be decoupled from the condition for the best image irradiance, and the areas of the best contrast and the highest resolution do not coincide with places along the optical axis where the recorded image is the brightest.

\textbf{Third,} in nonparaxial single-pixel THz inspection of an object involving structured illumination and image retrieval, for each given set of diffractive optical elements, unique rational design rules can be determined numerically via the Rayleigh-Sommerfeld diffraction integral. The quality of the recorded image depends not only on the distance $z_1$ from the illuminating element but also on the distance $z_2$ to the image retrieving element. These locations are particular for each combination of the THz light engineering/collecting diffractive elements and define the conditions for the best contrast, resolution, depth of field and focus, and brightness of the recorded image.

\textbf{Fourth,} in nonparaxial single-pixel THz imaging, the light illumination engineering is not the single condition for improving quality of \ac{THz} images -- conventional Gaussian illumination in combination with structured image retrieval using Bessel axicon and Airy zone plate contributes to the enhancements to the compact THz imaging setup in terms of contrast, resolution, and depth of field.


Lastly, we did find an indication that the optimal distances between the illuminating element and the object ($z_1$) and between the object and the detecting element ($z_2$) are functionally related in a unique way determined by the choice of the diffractive optical elements in the illumination and detection parts. An artificially introduced nonoptimal position $z_2$ of the detecting element in the setups used can be counteracted by adjusting the position $z_1$ of the sample with respect to the illuminating element. This behavior indicates that the condition for the best resolution implies a functional dependency $f(z_1,z_2)$ on those particular positions to illuminate and detect the object. Studies are underway on the determination of these relationships. 


Therefore, the presented findings open a ground-breaking pathway for the advancement of a new generation of compact nonparaxial THz imaging systems employing engineered and structured light in illumination or image retrieval schemes. In particular, they can play an essential role in the further development of miniature THz imaging systems that rely on the integration of optical components directly on a semiconductor chip. As a consequence, it will stimulate novel applications that enrich the scope of a rapidly evolving field of advanced THz technology \cite{Leitenstorfer2023}. \\

\section{Supporting Information} 

Animation to illustrate optical nonparaxiality in THz imaging is given as a separate file.
Other Supporting Information is available from the corresponding author upon reasonable request.\\

\section{Acknowledgements} 

This research has received funding from the Research Council of Lithuania (LMTLT): The experimental part of the research was supported via agreement No [S-MIP-22-76], theoretical investigation -- via agreement of agreement No [S-MIP-23-71].

\section{Conflict of Interest}

The authors declare no conflict of interest.\\

\section{Data Availability Statement}

The data supporting the presented findings in this study are available from the corresponding author upon reasonable request.

\section{Keywords}

Terahertz imaging, light engineering, nonparaxial optics, single-pixel imaging, silicon diffractive optics. 

\newpage
\providecommand{\noopsort}[1]{}\providecommand{\singleletter}[1]{#1}%
.

\end{document}